\documentclass[twocolumn,tighten]{aastex63}

\usepackage{color}

\def\om{\Omega_m}
\def\s8{\sigma_8}
\def\omb{\Omega_b}

\def\mqmsf{M_{\ast,q}/M_{\ast,\tiny sf}}
\def\ml{{\rm M}_h/{\rm L_{cen}}}
\def\fq{f_q}
\def\vmax{V_{\rm max}}

\def\cgal{c_{\rm gal}}
\def\magr{M_r-5\log h}
\def\lcen{L_{\rm cen}}

\def\dn{{\rm D}_n4000}
\def\dcrit{{\rm D_{crit}}}

\def\lgrp{L_{\rm grp}}
\def\psat{P_{\rm sat}}

\def\wp{w_p(r_p)}
\def\wchi{w_\chi}

\def\wcen{w_{\rm cen}}
\def\wcenc{w_{\rm cen,c}}
\def\wcenr{w_{\rm cen,q}}
\def\wcenb{w_{\rm cen,sf}}

\def\lgal{L_{\rm gal}}
\def\lsolhh{h^{-2}{\rm L}_\odot}

\def\fsat{f_{\rm sat}}
\def\msol{$M$_\odot}
\def\hmsol{h^{-1}{\rm M}_\odot}
\def\mgal{M_\ast}
\def\mhalo{M_h}
\def\slogl{\sigma_{\log L}}
\def\slogm{\sigma_{\log M_\ast}}
\def\lsat{L_{\rm sat}}
\def\lsatred{L_{\rm sat}^{\rm q}}
\def\lsatblue{L_{\rm sat}^{\rm sf}}

\def\hkpc{h^{-1}{\rm kpc}}
\def\hmpc{h^{-1}{\rm Mpc}}

\def\bsat{B_{\rm sat}}
\def\bsatc{B_{\rm sat,c}}
\def\bsatred{B_{\rm sat,q}}

\def\wred{w_{\rm red}}
\def\wblue{w_{\rm blue}}

\def\ltot{{L}_{\rm tot}}
\def\lsatchibar{\tilde{L}_{\rm sat}(\chi)}

\submitjournal{ApJ}

\shorttitle{Self-Calibrating Group Finder}
\shortauthors{Tinker}

\begin{document}

\title{A Self-Calibrating Halo-Based Group Finder: Application to SDSS}

\correspondingauthor{Jeremy L. Tinker}
\email{jeremy.tinker@nyu.edu}

\author[0000-0003-3578-6149]{Jeremy L. Tinker}
\affiliation{Center for Cosmology and Particle Physics, Department of
  Physics, New York University, New York, USA, 10003}

\begin{abstract}

  We apply a new galaxy group finder to the Main Galaxy Sample of the
  SDSS. This algorithm introduces new freedom to assign halos to
  galaxies that is self-calibrated by comparing the catalog to
  complementary data. These include galaxy clustering data and
  measurements of the total satellite luminosity from deep imaging
  data. We present constraints on the galaxy-halo connection for
  star-forming and quiescent populations. The results of the
  self-calibrated group catalog differ in several key ways from
  previous group catalogs and halo occupation analyses. The transition
  halo mass scale, where half of halos contain quiescent central
  galaxies, is at $\mhalo\sim 10^{12.4}$ $\hmsol$, significantly
  higher than other constraints.  Additionally, the width of the
  transition from predominantly star-forming halos to quiescent halos
  occurs over a narrower range in halo mass. Quiescent central
  galaxies in low-mass halos are significantly more massive than
  star-forming centrals at the same halo mass, but this difference
  reverses above the transition halo mass. We find that the scatter in
  $\log\mgal$ at fixed $\mhalo$ is $\sim 0.2$ dex for massive halos,
  in agreement with previous estimates, but rises sharply at lower
  halo masses. The halo masses assigned by the group catalog are in
  good agreement with weak lensing estimates for star-forming and
  quiescent central galaxies. We discuss possible improvements to the
  algorithm made clear by this first application to data. The group
  catalog is made publicly available.

\end{abstract}

\keywords{galaxies---groups: galaxies---halos}

\section{Introduction} \label{sec:intro}

One of the most striking features of the low-redshift galaxy population
is the bimodality of galaxy properties. Galaxies can generally be
categorized by their observed properties into two populations: the
blue cloud and the red sequence (e.g., \citealt{baldry_etal:04,
  baldry_etal:06, balogh_etal:04, blanton_moustakas:09}). Blue cloud
galaxies are actively star-forming, yielding blue colors and younger
stellar populations as probed by their 4000-Angstrom break
(\citealt{kauffmann_etal:03, brinchmann_etal:04}). Red sequence
galaxies have little to no star formation, resulting in red colors and
large 4000-\AA\ breaks in their spectra. Additionally, morphological
properties correlate with these classifications: at fixed stellar
mass, red galaxies are smaller, more concentrated, thus have higher
velocity dispersions and stellar surface mass densities
(\citealt{kauffmann_etal:03, blanton_etal:03cmd,
  blanton_moustakas:09}). Galaxy light profiles, quantified by
Sersic-index $n$, exhibit a unimodal distribution, but the fraction
of galaxies on the red sequence increases monotonically with
Sersic-$n$ (\citealt{blanton_etal:03cmd}). The fraction of galaxies
within each mode of the color distribution is highly dependent on the
large-scale environment around the galaxy (e.g.,
\citealt{baldry_etal:04}).

There is another galaxy property---though not an observable
quantity---that has a strong influence on this bimodality: whether or
not a galaxy is the central galaxy within its dark matter halo, or a
satellite galaxy orbiting within the gravitational potential of a
larger host halo. Galaxy group finders have been the most effective
tool for quantifying the impact of the central-satellite dichotomy on
the characteristics of the galaxy population (see, e.g.,
\citealt{weinmann_etal:06, blanton_berlind:07, yang_etal:08, yang_etal:09,
  peng_etal:12, wetzel_etal:12, wetzel_etal:13,
  tinker_etal:17_p1}). Among other results, these studies have
demonstrated that correlations between galaxy properties and
environment can be largely explained by simply separating galaxies
into these two classes of halo occupation: central and satellite
(\citealt{blanton_berlind:07, peng_etal:12, tinker_etal:17_p1,
  tinker_etal:18_p2, tinker:17_paris}).

These successes of galaxy group finders have also highlighted the
need for improvements in algorithms used. \cite{campbell_etal:15}
demonstrated that the halo masses estimated for groups, especially
when separating the central galaxies into star-forming and quiescent
populations, can be significantly biased. In Paper I
(\citealt{tinker:20_nextgen}), we presented a new galaxy group finder,
based on the halo-based algorithm of \cite{yang_etal:05}, that
ameliorates this issue, as well as improving on purity and
completeness of central and satellite galaxy classifications when
separating galaxies into their bimodal populations. The main
advancement of the group finding algorithm in Paper I is to introduce
new freedom into the model to account for possible unknown differences
in galaxy-halo connection for star-forming and quiescent galaxies, and
then self-calibrate these new degrees of freedom by comparing the
predictions of the group catalog to various observational statistics,
including galaxy two-point clustering and cross-correlations with
faint imaging galaxies around spectroscopic central galaxies.

Understanding how the galaxy-halo connection differs for galaxies on
opposite side of the bimodality is a key question in galaxy
formation. What role does the dark sector play in building up the red
sequence and making galaxies `red-and-dead?' Unfortunately, different
studies reach disparate conclusions. In a recent review,
\cite{wechsler_tinker:18} compiled observational constraints of the
galaxy-halo connection for star-forming and quiescent galaxy
population. The studies used halo occupation techniques to model
galaxy mass functions, clustering, and (for some) galaxy-galaxy
lensing, from the SDSS Main Galaxy Sample (MGS; \citealt{york_etal:00,
  strauss_etal:02}). At a given halo mass, $\mhalo$, these models
constrain the relative stellar masses, $\mgal$, of star-forming and
quiescent central galaxies. Some models predict quiescent galaxies are
more massive, some predict that star-forming centrals are more
massive, and others are consistent with the galaxies having the same
stellar mass. Each of these scenarios have different implications for
the relative important of the dark halo in galaxy quenching, and which
aspect of the dark halo---the halo mass or its formation history---is
determinative.

Galaxy group catalogs can help resolve the discrepancy between various
methods of quantifying the galaxy-halo connection. Unfortunately, they
require datasets that are both large-volume and highly complete, both
in terms of their target selection and in terms of redshift
completeness. In the near future, such datasets will be expanded by
large factors. The Bright Galaxy Survey of the Dark Energy
Spectroscopic Instrument survey (DESI-BGS; \citealt{desi_fdr}), will
produce an MGS-like sample of galaxies but larger by nearly a factor
of 20 both in terms of volume and number of spectra. The WAVES project
(\citealt{waves}) will produce both wide and deep surveys that will be
complementary to the BGS sample.  The MGS is an ideal testbed for
developing the next generation of galaxy group finders in advance of
upcoming data. The goal of this paper is both to understand the
galaxy-halo connection in the local universe, but also to
assess the efficacy of the self-calibrated algorithm with a
representative data sample.

This paper is organizes as follows: in \S 2 we present the data
utilized, including the measurements incorporated into the
self-calibration process. In \S 3, we review the self-calibrated
group-finding algorithm presented in Paper I, including
enhancements for analyzing the full MGS sample. In \S 4 we present our
results. The results will focus on the galaxy-halo connections for our
definitions of star-forming and quiescent, but in this section we will
also compare to results from myriad previous studies. In \S 5 we
discuss the results, both in terms of galaxy formation and in terms of
assessing the current algorithm and identifying aspects the require
improvement. As with Paper I, our results are based in part on the
Bolshoi-Planck simulation (\citealt{klypin_etal:16}). For
distance-redshift calculations, as well as theoretical calculations of
the halo statistics, we use the same cosmology as in that simulation:
($\om$, $\s8$, $\omb$, $n$, $h$)=(0.307, 0.82, 0.048, 0.96, 0.7). In
this work, we define dark matter halos as having a mean interior
density 200 times the background density. We also define halos as
being distinct---i.e., they do not exist within the radii of a larger halo.

\begin{figure*}
  \epsscale{1.0}
  \hspace{-0.4cm}
  \plotone{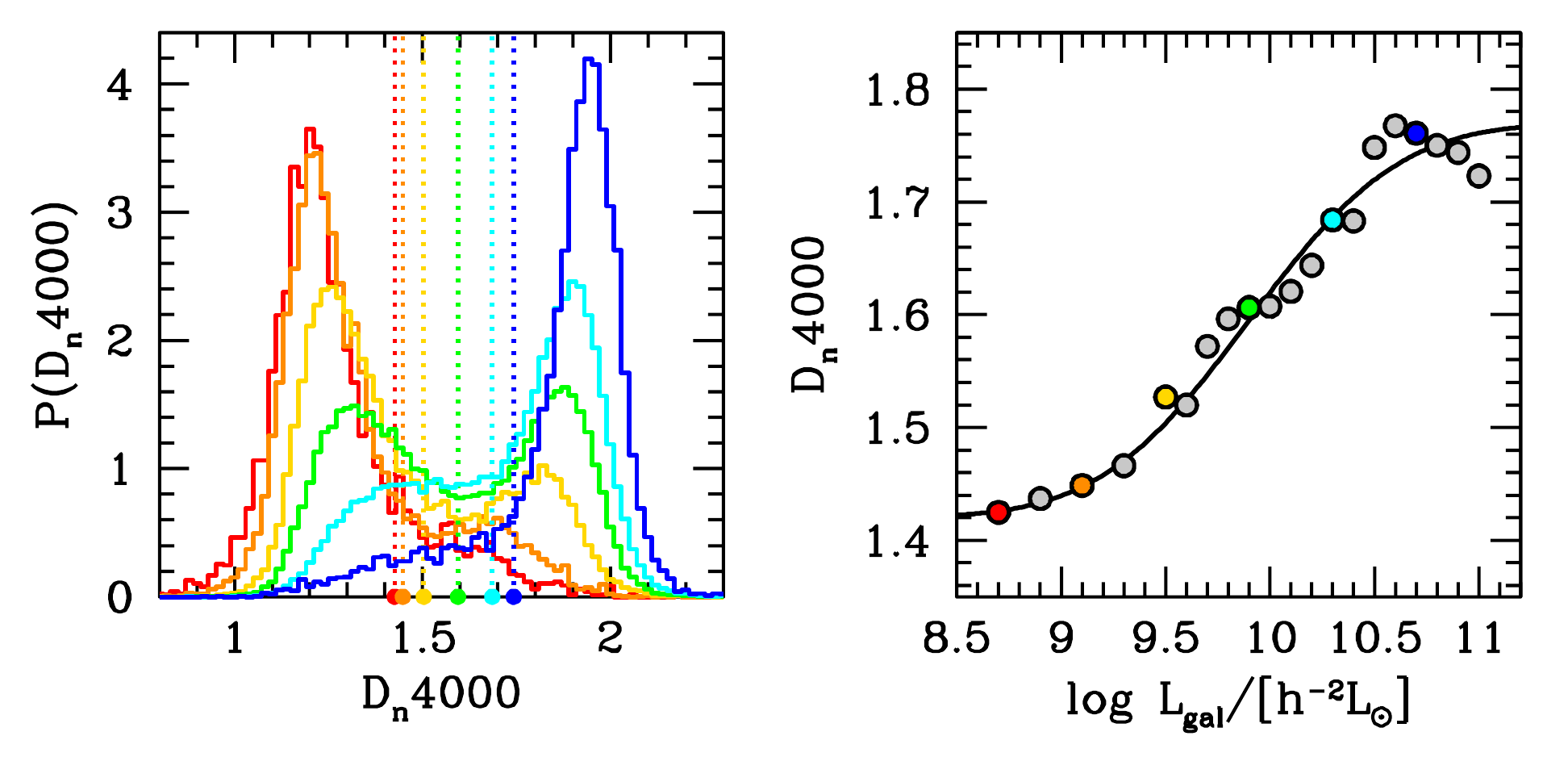}
  \vspace{-0.4cm}
  \caption{\label{f.dncut_gmm} {\it Left Panel:} Distributions of
    $\dn$ in 6 different bins of galaxy luminosity. The colors of each
    histogram correspond to the colored points in the right-hand
    panel, which indicates the value of $\lgal$ for each bin. The
    vertical dotted line ending in the filled circle on the $x$-axis
    is the break-point between the star-forming and quiescent
    populations derived from the Gaussian Mixture Modeling analysis
    described in the text. {\it Right PaneL:} The values of $\dn$
    where the two Gaussians in the GMM modeling have the same value,
    which we use as the population break point. The individual points
    represent the GMM results in each $\lgal$ bin, while the smooth
    curve is the fitting function used in the analysis. The vertical
    lines in the left-hand panel correspond to the values from the
    fitting function. }
\end{figure*}

\begin{figure}
  \epsscale{1.2}
  \hspace{-0.4cm}
  \plotone{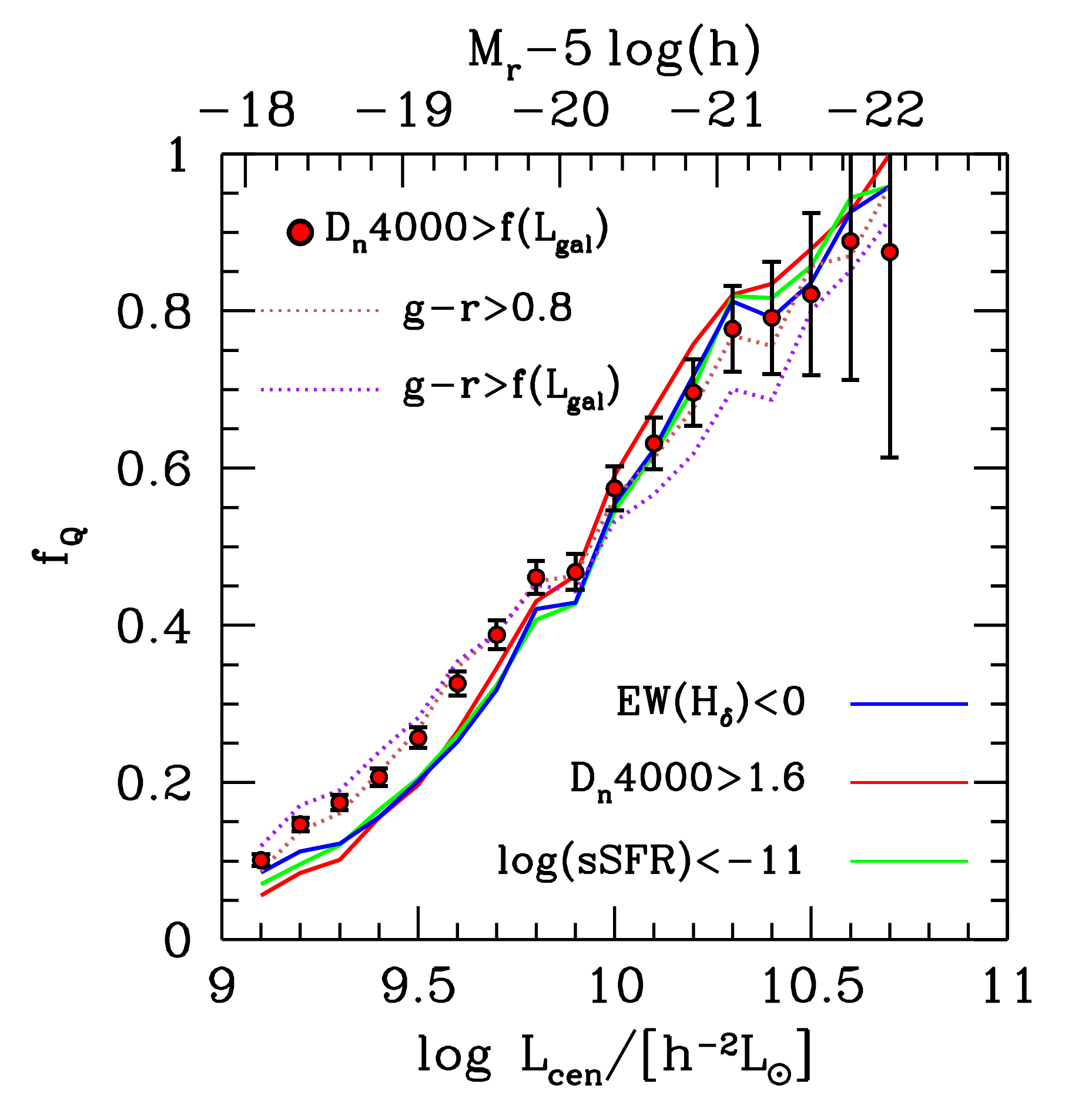}
  \vspace{-0.8cm}
  \caption{\label{f.fqcen_comp} Comparison of the quiescent fraction of
    central galaxies for various definitions of quiescent in the
    literature. All results are applied to the volume-limited group
    catalogs of \cite{tinker_etal:11}. The points with error bars
    represent the GMM split shown in Figure \ref{f.dncut_gmm}. The
    dotted lines indicate splits based on $g-r$ colors; the purple
    dotted line is the `tilted' cut from \cite{zehavi_etal:11} while
    the orange dotted curve represented the constant cut used in
    \cite{mandelbaum_etal:16}. The solid curves show splits based on
    spectroscopic quantities: the equivalent width of the H$_\delta$
    emission line, the 4000-Angstrom break, and the specific
    star-formation rate, sSFR. }
\end{figure}

\section{Data}
\label{s.data}

The data we use come from two main sources: SDSS MGS, and the
DESI Legacy Imaging Surveys (DLIS; \citealt{legacy_surveys}). For the
MGS, we use {\tt dr72bright34} sample from the NYU Value-Added Galaxy
Catalog (\citealt{blanton_etal:05_vagc}), which includes a total
sample of 559,028 spectroscopic galaxies\footnote{This includes the
  $\sim 7\%$ of galaxies that did not have a fiber assigned to them
  because of fiber collisions. We use the nearest-neighbor redshift
  assignment within the VAGC sample for distance and redshift. For
  properties derived from the spectroscopy, such as $\dn$, we assign
  each collided galaxy the properties of its nearest neighbor within
  color-magnitude space.}. Derived properties, specifically the galaxy
stellar masses, $\mgal$, and 4000-\AA\ breaks, $\dn$, come from the
MPA-JHU reductions, which are available with the SDSS DR8 public data
release and described in \cite{brinchmann_etal:04}. Although the
group-finding process is based on luminosity, we will examine the
catalog results as a function of stellar mass as well. Our fiducial
stellar masses are the PCA-based stellar masses of
\cite{chen_etal:12}, which were used in \cite{alpaslan_tinker:20}.

For all $\lsat$ measurements, we use DLIS DR6 and DR7. DR6 represents
northern imaging, Dec$>30^\circ$, from the Mayall and Bok telescopes
(\citealt{bass_mzls}), while the area below that declination is imaged by
DECam (\citealt{decam}). Together, these data cover roughly 75\% of the SDSS
footprint. The $\lsat$ measurements used in this analysis are
presented in \cite{tinker_etal:19_lsat} and
\cite{alpaslan_tinker:20}.

\begin{figure*}
  \epsscale{1.2}
  \hspace{-0.4cm}
  \plotone{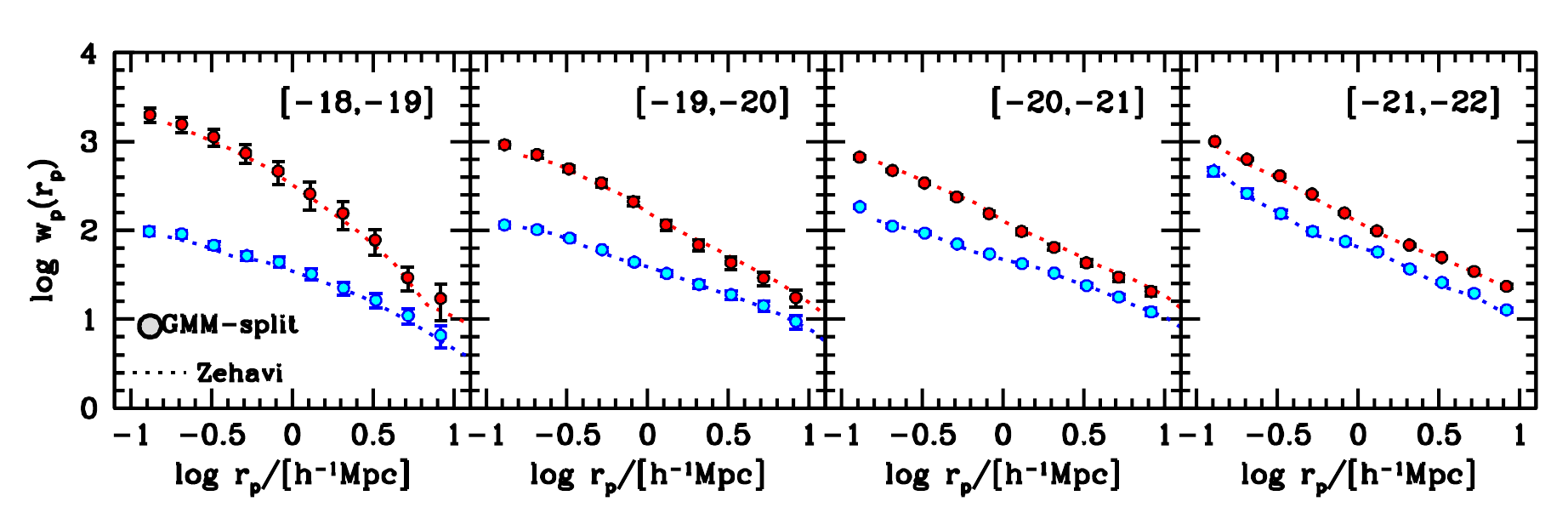}
  \vspace{-0.8cm}
  \caption{\label{f.wp} Projected correlation function in
    volume-limited samples within the full flux-limited MGS. The
    magnitude bin is indicated in the top right corner of each
    panel. Blue and red symbols indicate star-forming and quiescent
    galaxy samples, as constructed by the GMM analysis in Figure
    \ref{f.dncut_gmm}. These data are included in the self-calibration
    of the group finder. The dotted lines in the first three magnitude
    bins show the results from \cite{zehavi_etal:11}. There is no
    measurement from Zehavi for the brightest bin, so in the [-21,-22]
    bin the dotted lines show the results with a constant $\dn>1.6$
    split between star-forming and quiescence.}
\end{figure*}

\begin{figure}
  \epsscale{1.2}
  \hspace{-0.4cm}
  \plotone{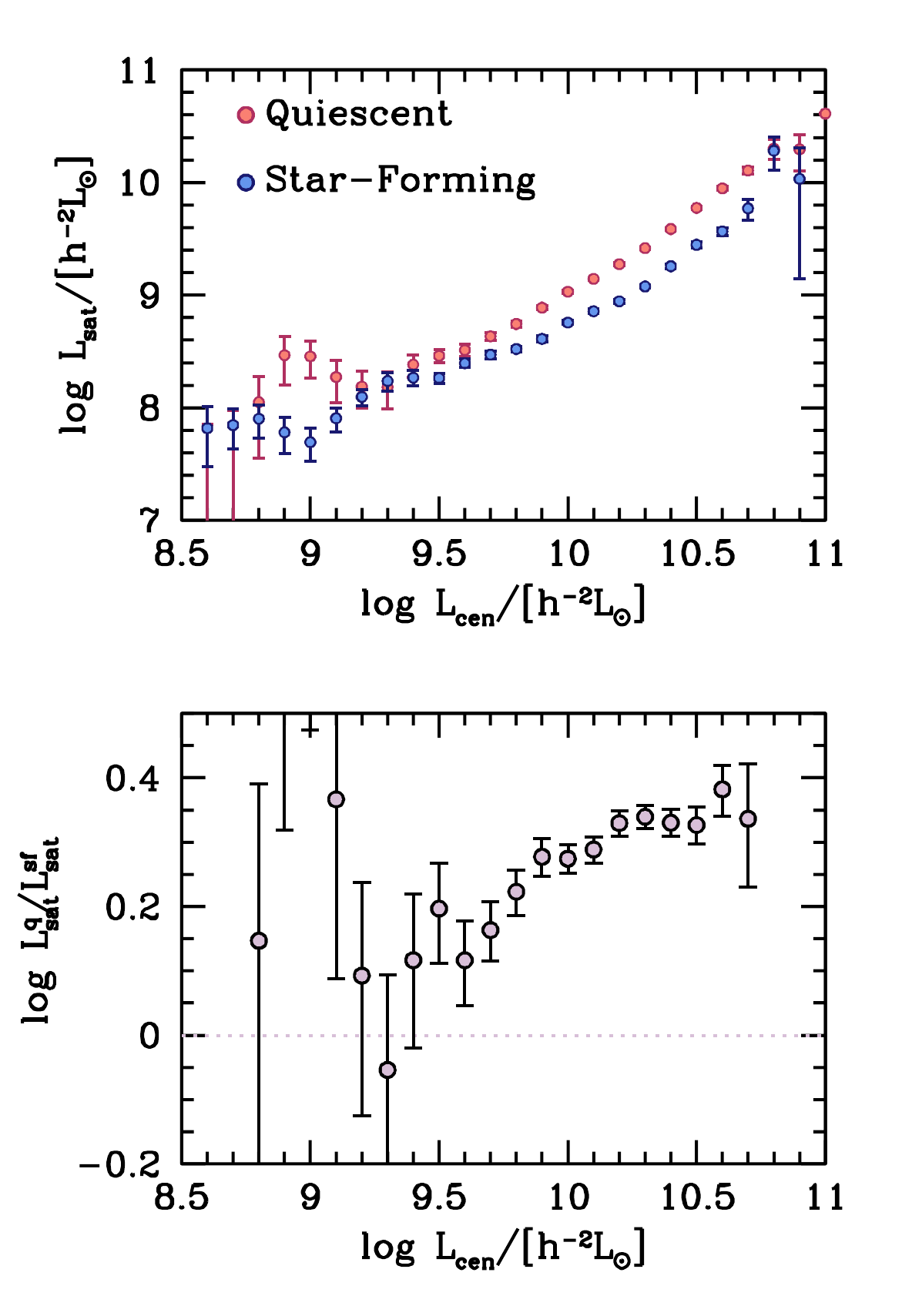}
  \vspace{-0.8cm}
  \caption{\label{f.lsat} {\it Top Panel:} $\lsat$ as a function of
    $\lcen$ for star-forming and quiescent subsamples of
    galaxies. Error bars are from the bootstrap method, sampling from
    the population of central galaxies. {\it Bottom Panel:} $\lsat$
    ratio between quiescent and star-forming galaxies at fixed
    $\lcen$. These data are the ones included in the self-calibration
    of the group finder.  }
\end{figure}

\begin{figure*}
  \epsscale{1.2}
  \hspace{-0.4cm}
  \plotone{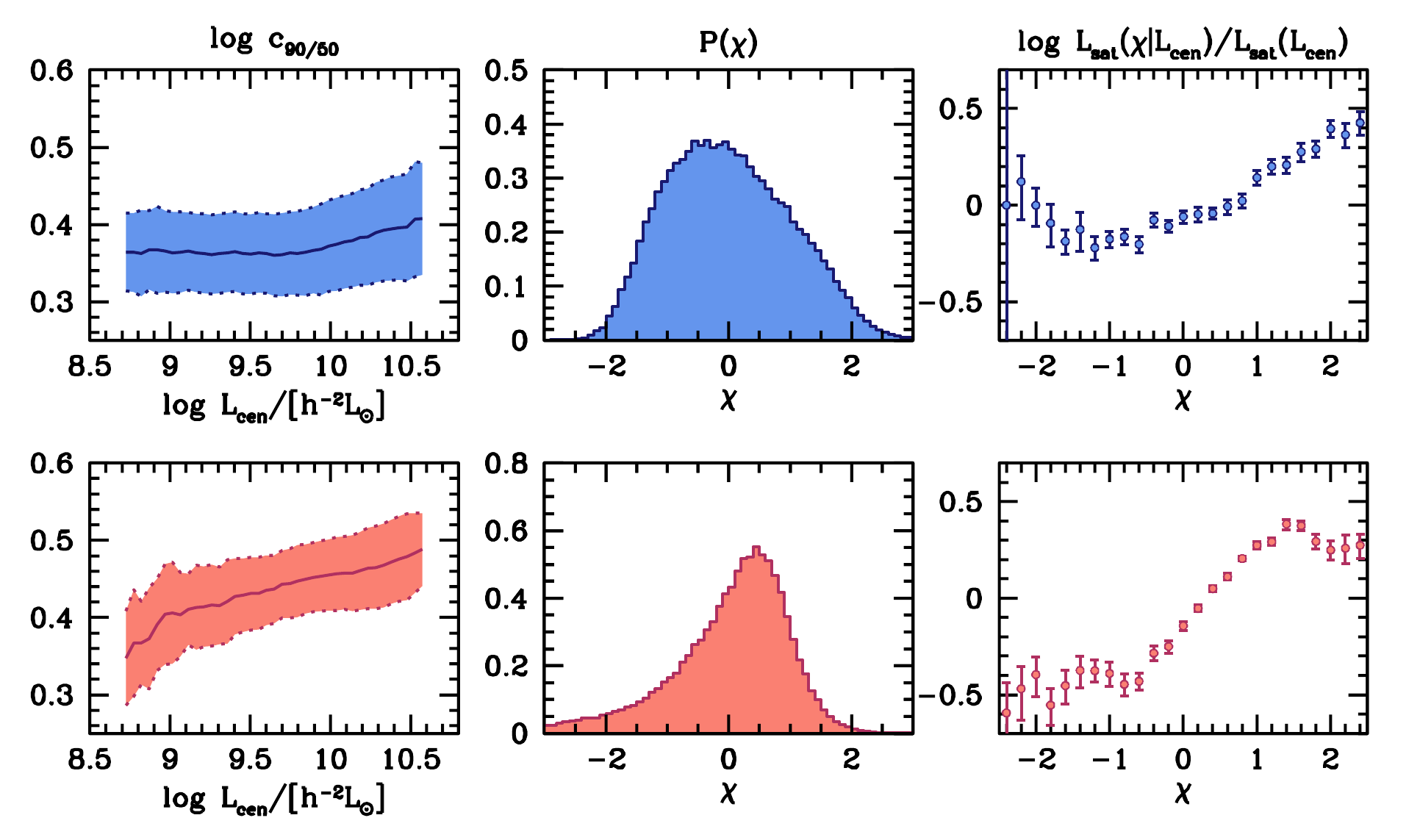}
  \vspace{-0.8cm}
  \caption{\label{f.con_propx} Incorporating the dependence of $\lsat$
    on galaxy concentration, $\cgal$, into the self-calibration
    method. The top row shows results for star-forming central
    galaxies, while the bottom row shows the results for quiescent
    central galaxies. {\it Left Column:} The correlation between
    $\log\cgal$ and $\lcen$. The solid curve shows the mean value,
    while the shaded region shows the $1$-$\sigma$ range of values. {\it
      Middle Column:} The distribution of $\chi$, defined as
    $(c-\bar{c})/\sigma_c$, where $\bar{c}$ is taken from the solid
    curve the left column, and $\sigma_c$ is the shaded region from
    those panels. {\it Right Column:} Correlation between $\lsat$ and
    $\chi$. Here, $\lsat$ for each galaxy is normalized by the mean
    value of $\lsat$ at that value of $\lcen$ for the given
    subsample. These data are incorporated into the self-calibration
    of the group finder. }
\end{figure*}

\subsection{Defining Quiescent and Star-Forming Samples}
\label{s.data_sfq}
  
In order to separate galaxies into quiescent and star-forming samples,
we must first define the break-point between these two
classifications. A number of definitions have been used across the
field: a single broadband color, color-color diagrams, star formation
rates, emission line equivalent widths, and $\dn$. Additionally, the
division between star-forming and quiescent can be a constant, or a
function of galaxy luminosity or stellar mass. 

In this paper, we use $\dn$ to separate the two populations of
galaxies. This quantity is significantly less sensitive to dust than
broadband colors, and any aperture bias is minimal---for galaxies
below the knee in the luminosity function, up to a quarter of those
classified as red by the $g-r$ color criterion are star-forming
galaxies reddened by dust (\citealt{zhu_etal:11}). $\dn$ identifies a
significant fraction of low-luminosity galaxies that are classified as
red by dust attenuation, but are intrinsically star-forming
(\citealt{tinker_etal:11}).  Although the fiber aperture of SDSS
spectra only subtends the central part of the galaxy, misidentifying a
blue galaxy as quiescent through $\dn$ is relatively rare, and usually
occurs for very low redshift objects (\citealt{geha_etal:12}).

It is convenient to set the threshold between star-forming and
quiescent samples as a constant value of $\dn$ or $g-r$, but the red
sequence has a `tilt,' such that brighter quiescent objects have
redder colors and larger $\dn$ values. This has motivated many studies
to have dividing lines that vary with galaxy mass or luminosity
(e.g., \citealt{li_etal:06, zehavi_etal:11}) To account for this, we use
Gaussian-mixture modeling (GMM) to determine the two populations as a
function of $\lgal$. Figure \ref{f.dncut_gmm} shows the distribution
of $\dn$ for multiple bins in $\log\lgal$. In each bin, we fit a
two-Gaussian model to the distribution. We set the $\dn$ value at which
the two Gaussians cross as the threshold separating star-forming and
quiescent populations. The vertical lines in this panel show this
treshold value.

The right-hand panel in Figure \ref{f.dncut_gmm} shows the results of
the GMM in each $\lgal$ bin. The solid curve represents a fit to these
results, of the form

\begin{equation}
  \label{e.fqbreak}
  \dcrit = 1.42 + \frac{0.35}{2}\left[1+{\rm erf}\left(\frac{\log\lgal-9.9}{0.8}\right)\right],
\end{equation}

\noindent where erf is the error function. We use
Eq.~(\ref{e.fqbreak}) to separate star-forming and quiescent galaxy
samples.

What impact do these choices on the quiescent fractions of galaxies?
For the purposes determining the LHMR of galaxies, it is the quiescent
fraction of central galaxies that is most important. The relative
numbers of star-forming and quiescent galaxies are a factor in
determining the halo masses, since each halo must have one central
galaxy.  Figure \ref{f.fqcen_comp} compares $\fq$ for multiple
definitions of quiescence found in the literature. The sample of
galaxies is taken from the volume-limited group catalog for
$\magr<-18$ using the standard halo-based group finder of
\cite{tinker_etal:11}. This figure is for comparative purposes only,
but we note that reproducing this figure using the final results of
this paper yield negligible results in the differences between the
definitions.

The GMM method of this paper is shown with the filled circles. Splits
based on $g-r$ color are indicated with dotted curves, while common
methods based on spectroscopic quantities are shown with the solid
curves. The two color-based cuts are taken from
\cite{mandelbaum_etal:16} and \cite{zehavi_etal:11}. The constant
$g-r>0.8$ is utilized by the former, while the latter study uses the
tilted threshold that varies linearly with $\magr$. All of the results
show qualitatively similar results, with the photometric splits
yielding slightly higher $\fq$ at low luminosities. This is likely
due to the presence of dust in the central galaxies. Because the GMM
results move the $\dcrit$ threshold lower for fainter galaxies, our
GMM split is most comparable to the constant $g-r$ threshold.

\subsection{Projected Galaxy Clustering}

Now that we have established the separation between star-forming and
quiescent galaxies, we turn to measurements of their
clustering. Figure \ref{f.wp} shows the projected correlation function
in four bins of magnitude. For each magnitude bin, we have constructed
volume-limites samples within which to measure the clustering. Error
bars are calculated using the jackknife technique, separating the
distribution of galaxies and randoms into 25 roughly equal-area
regions in the plane of the sky. The clustering results here are in
agreement with previous measurements at low-redshift galaxy
clustering, with the red galaxies having higher clustering amplitude
at all scales and for all magnitudes.

Figure \ref{f.wp} also compares our measurements to those of
\cite{zehavi_etal:11}. Zehavi uses the DR7 SDSS MGS, as
well as the same binning in magnitude. The only difference is the
star-forming/quiescent threshold. Figure \ref{f.fqcen_comp} demonstrated
that the difference in central galaxies is minimal, and this is
reflected in the clustering measurements. 

\subsection{Total Satellite Luminosity}

As presented in Paper I, one of the key ingredients in the
self-calibration method is the inclusion of measurements of total
satellite luminosity, $\lsat$. A full explanation of the method is
presented in \cite{tinker_etal:19}. In brief, $\lsat$ measures the
total amount of $r$-band luminosity around spectroscopic central
galaxies using deep imaging data from the DLIS. Because there is no
redshift information in the imaging galaxies, $\lsat$ is measured by
first stacking a set of central galaxies in a given bin and
subtracting a background value of imaging galaxies. For all
spectroscopic galaxies, $\lsat$ is measured within fixed comoving
apertures of 50 $\hkpc$.

Paper I demonstrated that these data provide significant constraining
power on the on the relative halo masses of star-forming and quiescent
central galaxies at fixed $\lcen$. Using only the satellites available
in the spectroscopic catalog is not sufficient to determine if these
two subsamples of galaxies live in different halos at fixed
luminosity.

Figure \ref{f.lsat} shows the measurements of $\lsat$ for star-forming
and quiescent central galaxies that are incorporated into the
self-calibrated group finder. The top panel shows the raw measurements
of $\lsat$ as a function of $\lcen$. The bottom panel shows the
actually data that the group finder is fit to: the relative
measurements at fixed $\lcen$, $\lsatred/\lsatblue$. The reason behind
using the relative values rather than the absolute values is that it
reduces systematic errors, both in terms of the measurements and the
theoretical modeling. Errors in the background subtraction, as well as
miscalibrations between the Legacy Surveys imaging data and the SDSS
imaging, which we use here to make our model predictions (which we
describe in more detail in \S \ref{s.methods} and Paper I). Additionally,
numerical resolution can impact the number of subhalos
(\citealt{vandenbosch_etal:18, vandenbosch_etal:18b}). This can lower
the predicated value of $\lsat$ from a simulation, but the effects
are reduced when calculating the ratio of satellite luminosities.

The bottom panel in Figure \ref{f.lsat} shows the $\lsatred/\lsatblue$
ratio. For bright central galaxies, quiescent central galaxies are
halos with roughly twice the satellite luminosity (within the
aperture). At lower luminosities, the data suggest that the ratio gets
closer to unity, but the large error bars make prevent any conclusions
from inspection of the measurements.


\subsection{Secondary Galaxy Properties}

In addition to measuring $\lsat$ as a function of luminosity, the key
result from \cite{alpaslan_tinker:20} is that most galaxy secondary
properties also carry information about their dark matter halos. At
fixed galaxy stellar mass, $\lsat$ correlates with properties such as
stellar velocity dispersion, galaxy size, and morphological
properties.

These correlations do not, by themselves, imply correlations between
galaxy properties and $\mhalo$. Halo formation history also impacts
the amount of subhalos within a host halo, thus correlating with the
amount of satellite luminosity. To break this degeneracy, the test
proposed by \cite{tinker_etal:19} is to measure the correlation
between the central galaxy secondary secondary property and the
large-scale environment. If the correlation between $\lsat$ and
secondary property is due to a correlation between halo formation
history and the property, the property will exhibit a correlation with
the large-scale environment. This is due to the correlation between
halo formation history and large-scale environment: older, more
concentrated halos form in high density regions of the cosmic web.

To increase our ability to properly assign halo masses to central
galaxies, we use galaxy concentration as the secondary
parameter. Concentration, $\cgal$, is defined as the ratio between the
radius that contains 90\% of the galaxy light to the half-light
radius. The are several reasons behind this choice: (1) $\cgal$ for
central galaxies shows no correlation between with large-scale
environment at fixed stellar mass, (2) $\cgal$ shows a correlation
with $\lsat$ that is roughly independent if $\mgal$
(\citealt{alpaslan_tinker:20}), and (3) $\cgal$ varies minimally with
galaxy luminosity.

As in Paper I, we transform the secondary property into a normalized
parameter, $\chi \equiv (\cgal-\bar{c})/\sigma_c$, where $\bar{c}$
and $\sigma_c$ are the mean and standard deviation of $\cgal$ and are continuous functions of $\lcen$. Figure
\ref{f.con_propx} shows $\bar{c}$ and $\sigma_c$ as a function of
$\lcen$ separately for star-forming and quiescent galaxies. The middle
panels show the distributions of $\chi$ for the two populations of
central galaxies.

We take the measurements of $\lsat$ for each galaxy and normalize it by
the expected value of $\lsat$ given the galaxy luminosity. This is
given by the mean $\lsat-\lcen$ relation, derived independently for
star-forming and quiescent central galaxies. This removes the overall trend of
$\lsat$ with luminosity, leaving only the residuals of $\lsat$ as a
function of $\chi$. Galaxies are binned by $\chi$ and in each bin the
mean value is calculated;

\begin{equation}
  \lsatchibar = \frac{1}{N}\sum_{i} \frac{{\lsat}_{,i}}{\lsat({\lcen}_{,i})},
\end{equation}

\noindent where $N$ is the number of galaxies in the bin of $\chi$,
and $i$ is the index of all galaxies within the bin of $\chi$.

The results for both star-forming and quiescent subsamples in the
right-hand column of Figure \ref{f.con_propx}. As expected from the
\cite{alpaslan_tinker:20} results, both of these galaxy subsamples
show significant correlations between $\chi$ and $\lsat$. But here the
measurement has significantly higher signal-to-noise given that all
galaxies are combined into a single measurement. For star-forming
galaxies, $\lsat$ is positively correlated with $\chi$, meaning
higher-concentrated galaxies live in halos with more satellite
luminosity, most likely due to higher halo masses. The values of
$\lsat$ vary by nearly 0.6 dex, or a factor of four. The correlation
for quiescent galaxies is even more extreme: varying by nearly 0.8 dex
in a smaller range of $\chi$, from -1 to +1. Outside of this range,
the trends plateau at their maximum and minimum values.

\section{Methods}
\label{s.methods}

A full outline of the self-calibration method is presented in Paper
I. Here we briefly review the algorithm, introducing the variables
that will be fit in the self-calibration process.

\subsection{Free parameters in the model}
\label{s.methods_params}

The halo-based group finder uses an initial estimate of the halo mass
around a central galaxy to determine the probability that neighboring
galaxies are satellites within the halo. Then the halo mass is
re-estimated by a rank-ordering of the halos by their total
luminosity, which is abundance-matched onto the halo mass
function. A neighbor is considered a satellite if $\psat>0.5$, where

\begin{equation}
  \label{e.psat}
  P_{\rm sat} = \left[1-\left(1+P_{\rm proj}P_{z}/\bsat\right)^{-1}\right],
\end{equation}

\noindent where $P_{\rm proj} $ and $P_{z}$ are the projected and
line-of-sight probabilities, respectively. $\bsat$ is a free parameter
setting the threshold probability. In the standard group-finder,
$\bsat=10$ independent of galaxy type or luminosity. In the
self-calibrated algorithm, $\bsat$ is a function of $\lgal$,
parameterized as

\begin{equation}
  \label{e.bsat}
  \bsatc = \beta_{0,c} + \beta_{L,c}(\log\lgal-9.5),
\end{equation}

\noindent where the class of the galaxy is indicated by $c=sf$ or
$q$ for star-forming and quiescent galaxies, respectively. $\bsat$
cannot be negative, thus we implement a minimum value of $\bsat$ of
0.01 when implementing the group finder.

Once all galaxies have been assigned to groups, halo masses are
assigned to each group using abundance matching. In the standard
implementation of the halo-based group finder, the total luminosity,
$\ltot$, (or total stellar mass, $M_{\ast\rm tot}$) of the group is
abundance matched onto the theoretical estimate of the host halo mass
function. In the self-calibrated model, we allow weight factors to be
placed on a group given certain properties of the group.

First, we put separate weight factors on groups with star-forming and
quiescent central galaxies. These weight factors are functions of
$\lcen$, such that

\begin{equation}
  \label{e.wcen}
  \log \wcenc =\frac{\omega_{0,c}}{2}\left[1+{\rm erf}\left(\frac{
      \log\lgal-\omega_{L,c}}{\sigma_{\omega, c}} \right)\right], 
\end{equation}

\noindent where $c$ indicates central galaxy class as in
Eq.~(\ref{e.bsat}). Eq.~(\ref{e.wcen}) allows our two classes of
galaxies to occupy different halos, even if the total group luminosity
is the same.

To incorporate the information available in the $\lsatchibar$ data, we
introduce weights based on a central galaxy's $\chi$ value, with

\begin{equation}
  \label{e.wchi}
  w_{\chi,c} = \exp\left[\frac{\chi}{\omega_{\chi,c,0}+\omega_{\chi,c,L}(\log\lgal-9.5)}\right],
\end{equation}

\noindent where the exponential form is motivated by the log-linear
dependence of $\lsat$ on most galaxy properties at fixed $\mgal$
(\citealt{alpaslan_tinker:20}), as well as the results of
$\lsatchibar$ in Figure \ref{f.con_propx}. Eq.~(\ref{e.wchi})
introduces two new free parameters over the model in Paper I, in which
the $w_\chi$ weights were independent of $\lgal$. This was sufficient
in the tests presented in Paper I because the mock galaxy
distributions were constructed with correlations between $\chi$ and
$\mhalo$ that were also independent of $\lgal$. In fitting the SDSS
data, we find statistically improves results with this additional
freedom.

This makes 14 total free parameters. These parameters are listed in
Table 1. We also list the 68\% confidence intervals resulting from the
fitting procedure we describe in \S \ref{s.methods_outline}.

When rank-ordering groups, we use the weighted total luminosity,
$L_{\rm grp} = \ltot\times \wcenc\times w_{\chi, c}$. We define
$\ltot$ as the total $r$-band luminosity of spectroscopic galaxies
within the group. This marks
another minor change from Paper I, in which the weight factors were
only applied to the central galaxy luminosity. Although the central
galaxy dominates to group luminosity for the vast majority of groups,
we find that the change yields an improved fit to the SDSS data.

\subsection{Adaptation for flux-limited samples}

Paper I demonstrated that the group finder efficiently identifies
groups in volume-limited samples. Thus, when applying the algorithm to
the flux-limited MGS sample, we treat the full sample as a series of
smaller-volume samples, each of which is much closer to a
volume-limited sample than the full sample.

Each sub-volume has a redshift thickness of 0.05. Within each
sub-volume, all galaxies above the flux limit are included. Thus, to
account for the change in the magnitude limit across the redshift
width of the bin, we use $1/\vmax$ weighting to calculate the number
density of groups, based on the $\vmax$ value of the central
galaxy. If $\vmax$ is larger than the volume of the of the upper
redshift limit of the bin, $\vmax$ is set to be that volume of the
upper limit. Thus, for the majority of groups in each redshift bin, $\vmax$
is equal to the volume of the bin itself.

To avoid discontinuities in how halos are assigned to galaxies, the
redshift bins are finely spaced by only $\Delta z=0.005$, and allowed
to overlap. Thus a given group contributes to the number density in 10
bins. When assigning halo mass to a given group, we use the number
density in the bin that is closest to the redshift of group's central
galaxy. The abundance-matching expression used to convert weighted
group luminosity to halo mas is

\begin{equation}
  \label{e.am_vmax}
\sum_{L=\lgrp}^{\infty}\frac{1}{\vmax} =
\int_{\mhalo}^\infty n_h(\mhalo^{\prime})d\mhalo^{\prime},
\end{equation}

\noindent where $n_h$ is the halo mass function
(\citealt{tinker_etal:08_mf}), and $\vmax$ is constrained by the
redshift bin to which Eq.~(\ref{e.am_vmax}) is applied, as discussed
above.

\subsection{Outline of the method}
\label{s.methods_outline}

To find the optimal set of parameters, we compare predictions from the
group catalog to our observational quantities. In order to make these
predictions, we use the group catalog produced by a given set of
parameters to populate halos in an N-body simulation. For this
analysis we use the Bolshoi Planck simulation (\citealt{klypin_etal:16}).

The procedure for implementing the self-calibrated group finder is as
follows:

\begin{itemize}
  \vspace{-0.2cm}
  \item Start with a set of values for the 14 free parameters of the
    model.
  \vspace{-0.2cm}
  \item Run the group finder with this set of parameters, iterating
    until the satellite fraction converges. This assigns $\mhalo$ to
    each group and classifies every galaxy as central or satellite.
    \vspace{-0.2cm}
  \item Measure the halo occupation distribution (HOD; see
    \citealt{wechsler_tinker:18}) within the resulting group catalog
    for each volume-limited bin for which $\wp$ is measured in the
    SDSS data.
  \item Populate the host halos of the Bolshoi-Planck N-body simulation with the HODs for
    each magnitude bin and measure the predicted clustering.
    \vspace{-0.2cm}
  \item Calculate the predicted value of $\lsatred/\lsatblue$ for the
    group catalog using tabulated values for $\lsat$ as a function of
    $\mhalo$.
  \vspace{-0.2cm}
  \item Calculate the predicted value of $\lsatchibar$ for the group
    catalog using the same tabulation of $\lsat$ as a function of
    $\mhalo$.
  \vspace{-0.2cm}
  \item Calculate the $\chi^2$ value for the model by comparing the
    grop catalog predictions to the SDSS data.
\end{itemize}

\noindent Following the outline above, for a given model, the total
$\chi^2$ is

\begin{equation}
  \label{e.chi2tot}
\chi^2_{\rm tot} = \sum_{\rm sf,q}\sum_{i=1}^{N_{\rm
    bins}}\chi^2_{w_p,i} + \chi^2_{Lsf/Lq} + \sum_{c=\rm sf,q}\chi^2_{L\chi,c}
\end{equation}

\noindent where the first term is the $\chi^2$ from the clustering
data, summing over the two subsamples and four magnitude bins, the second term
is from the $\lsatred/\lsatblue$ data, and the last term is the
$\lsatchibar$ data, summing over both subsamples. When calculating
$\chi$, we add in quadrature the errors from the observational data
with the sample variance expected from the Bolshoi Planck simulation
volume, which is 250 $\hmpc$ on a side. 

To find the best-fitting model, we use Powell's method to minimize
$\chi^2_{\rm tot}$. To find the confidence intervals on the
parameters, and thus our constraints on the halo masses assigned to
each central galaxy, we use {\tt emcee}, a public implementation of
the Markov Chain Monte Carlo method (\citealt{emcee}). 

\begin{figure*}
  \epsscale{1.2}
  \hspace{-0.4cm}
  \plotone{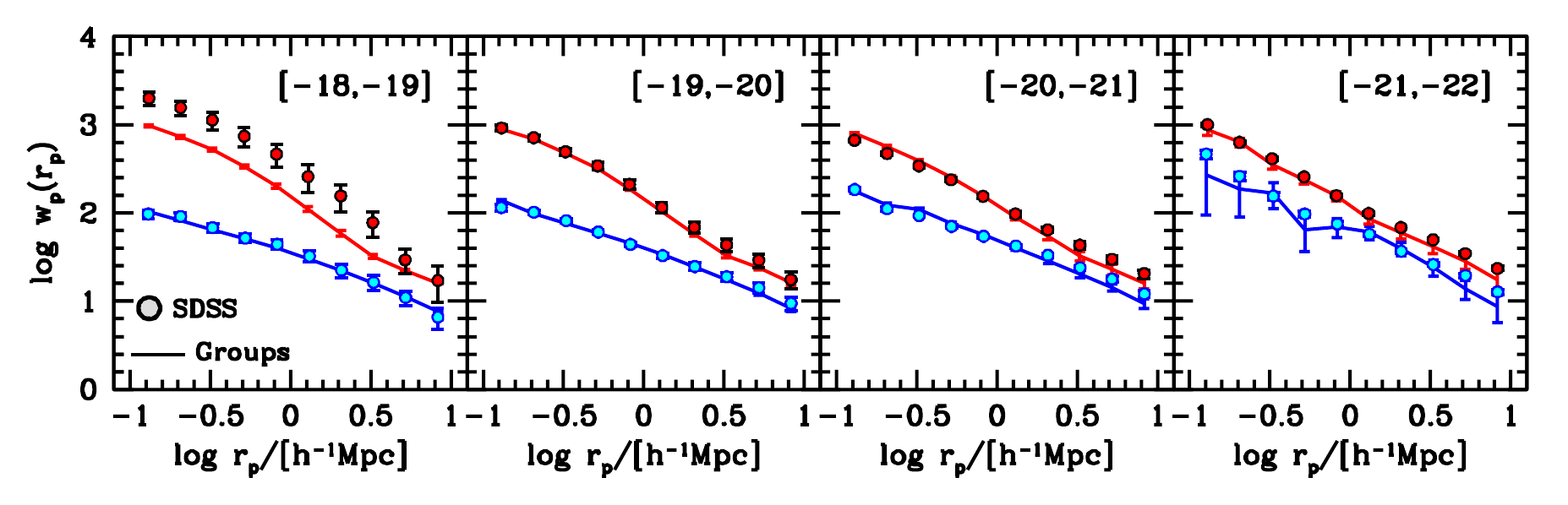}
  \vspace{-0.4cm}
  \caption{\label{f.wp_fit} Projected correlation functions yielded by
    the best-fit galaxy group catalog. Points with errors are the SDSS
    measurements from Figure \ref{f.wp}. The solid curves are the
    best-fit model. Red points and curves indicate quiescent samples,
    while blue points and curves represent star-forming samples. The
    errors on the group catalog predictions are from the simulation
    volume, while the errors on each SDSS measurement are derived from
    sample sizes that increase with the magnitude limit. Thus, for the
    faint bin the observational error dominates, while for the
    brightest bin the theoretical error dominates.}
\end{figure*}

\begin{figure*}
  \epsscale{1.}
  \hspace{-0.4cm}
  \plotone{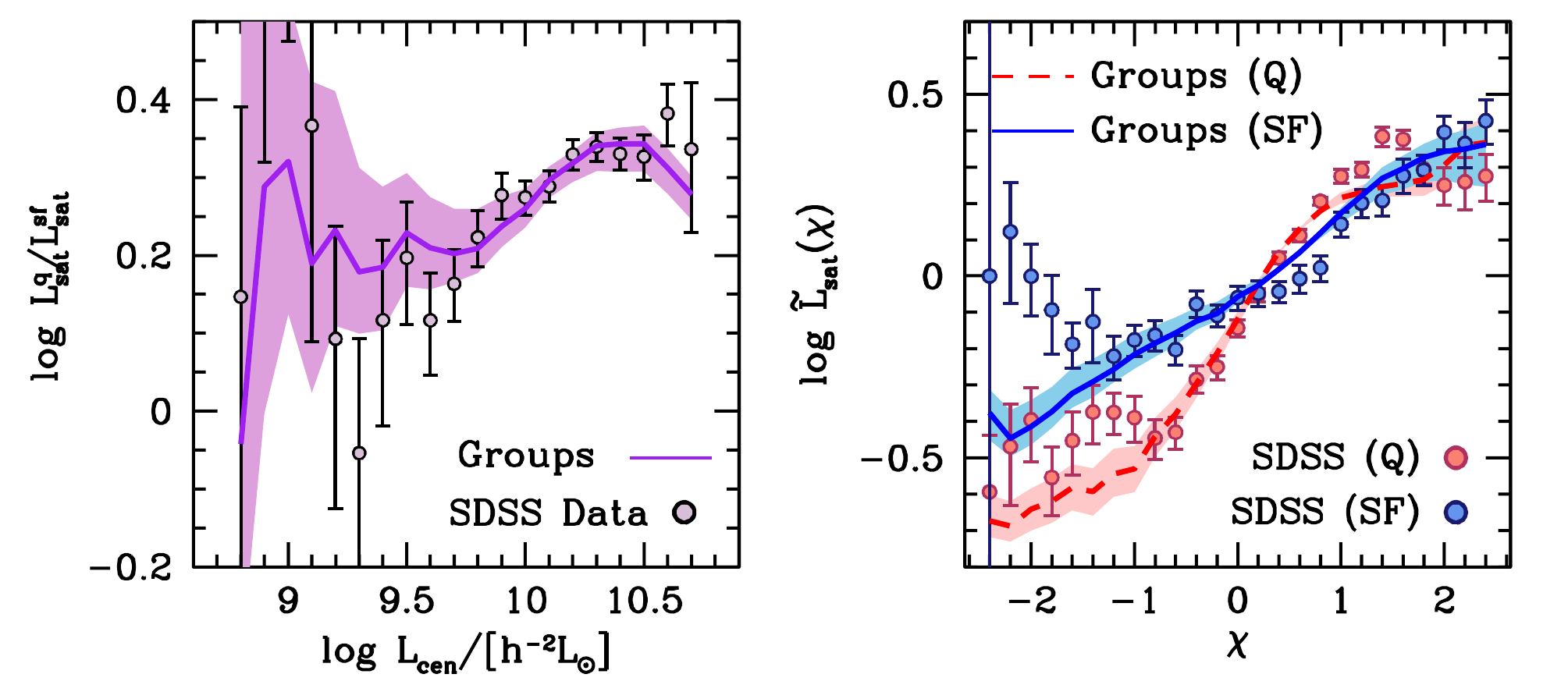}
  \vspace{-0.1cm}
  \caption{\label{f.lsat_propx} {\it Left Panel:} $\lsatred/\lsatblue$
    yielded by the best-fit group catalog. Points with error bars are
    the $\lsat$ SDSS measurements from Figure \ref{f.lsat}. The shaded
    region is the 95\% confidence interval on this quantity. {\it
      Right Panel:} $\lsatchibar$ produced by the best-fit group
    catalog. Points with error bars are the SDSS data from Figure
    \ref{f.con_propx}. Red symbols and curve represent quiescent
    central galaxies, while blue symbols and curves represent
    star-forming central galaxies. Shaded regions around the curves
    are 95\% confidence intervals.}
\end{figure*}

\section{Results}
\label{s.results}

Here we present the results of the group finding process. We focus
first on how well the group catalog can reproduce the measurements
from SDSS, and the constraints they yield on the free parameters of
the group finder. Next, we discuss the derived properties of the
galaxy population. These include the satellite fractions of galaxies,
the relations between halo mass and central galaxy
properties---specifically luminosity and stellar mass, and the scatter
between them. For many results, we will compare those derived from the
fiducial self-calibrated group catalog to results in which the
$\lsatchibar$ data have been removed from the analysis, and all
$\wchi$ weights are set to unity. We will refer to this catalog as
`no-$\chi$.' The primary results will be referred to as the fiducial
catalog. Where relevant, we will compare to complementary
analyses from parametric halo occupation models and previous group finders.

 \begin{deluxetable*}{ccccc}
\tablenum{1}
\tablecaption{Parameters of the Self-Calibrated Group Finder\label{tab:messier}}
\tablewidth{0pt}
\tablehead{
\colhead{Parameter} & \colhead{68\% low} & \colhead{Median} &
\colhead{68\% high} & \colhead{Best fit} }
\startdata
$\beta_{0,sf}$ & 11.79 & 13.52 & 15.17 & 12.49\\
$\beta_{L,sf}$ & -10.38 & -8.26 & -6.59 & -8.72\\
$\beta_{0,q}$ & -1.22 & -0.94 & -0.69 & -0.91 \\
$\beta_{L,q}$ & 7.80 & 10.50 & 12.76 & 10.36 \\
$\omega_{0,sf}$  & 13.34 & 17.64 & 22.97 & 16.84 \\
$\omega_{L,sf}$  & 12.59 & 13.11 & 13.69 & 13.03\\
$\sigma_{\omega,sf}$  & 1.93 & 2.39 & 2.83 & 2.39\\
$\omega_{0,q}$  & 2.04 & 2.67 & 3.37 & 2.95\\
$\omega_{L,q}$  & 12.06 & 13.02 & 14.22 & 12.63\\
$\sigma_{\omega,q}$  & 3.70 & 4.87 & 6.28 & 4.20\\
$\omega_{\chi,sf,0}$ & 2.45 & 2.79 & 3.30 & 2.68\\
$\omega_{\chi,sf,L}$ & 1.75 & 2.18 & 2.58 & 2.34\\
$\omega_{\chi,q,0}$ & 1.03 & 1.12 & 1.23 & 1.11 \\
$\omega_{\chi,q,L}$ &0.31 & 0.47 & 0.57 & 0.43\\
\enddata

\tablecomments{The first three numerical columns indicate the 68\% confidence
  regions on each parameter from the MCMC analysis, as well as the
  median value of each parameter. The final column indicates the
  values of each parameter in the model that yields the lowest
  $\chi^2$ value.}
\end{deluxetable*}

\subsection{Fits to the data}

To recap, we have data from three different observed quantities: the
projected correlation function, $\wp$, the ratio of $\lsat$ for
star-forming and quiescent central galaxies, and the dependence of
$\lsat$ on central galaxy concentration, separated into measurements
for star-forming and quiescent central galaxies. This yields a total
data vector of 148 data points. The best-fit group catalog yields a
$\chi^2$ of 165. For 14 degrees of freedom, this value yields a
$P(>$$\chi^2)$ of 0.04. So while the the model is a reasonable
description of the data, the random chance of obtaining that
$\chi^2$ value is roughly a
$2\sigma$ deviation. The places were the fit can be improved in future
iterations of the model are clear upon inspection of the fits
themselves. We will discuss these subsequently.

Figure \ref{f.wp_fit} shows the best-fit group catalog with the SDSS
measurements of $\wp$. At all luminosities and all scales, the group
catalog yields enhanced clustering relative to star-forming
galaxies. In most all magnitude bins, the group catalog fit is in
excellent agreement with the data. However, the quiescent galaxies in
the $\magr=[-18,-19]$ bin is clearly discrepant, significantly
underpredicting the clustering amplitude. Fort this one
bin\footnote{Although the fainter $\magr=[-17,-18]$ bin is not
  included in the fit, we note that the the group catalog also
  underpredicts the clustering amplitude of quiescent galaxies in this
  bin as well, while being in good agreement with the 
  clustering of the star-forming sample. The error bars on this sample are too large for it's
  inclusion in the fit to have a significant effect on the best-fit
  model.}, the extra freedom incorporated into the group finder is
still not sufficient to match the data. We will discuss this in more
detail in the next subsection., but this is the clearest place to
improve the $\chi^2$ of the model.

Figure \ref{f.lsat_propx} compares the group catalog to the different
$\lsat$ measurements. The left-hand panel presents
$\lsatred/\lsatblue$, where the shaded region around the best-fit
catalog is the 95\% confidence interval on this quantity. The catalog
yields a good fit to the data at all luminosities, but the constraints
on the $\lsat$ ratio widen significantly at low luminosities due to
the large errors on this statistic. At the brightest luminosities, the
turnover in the group catalog values is due to aperture for which
$\lsat$ is calculated. As halos get more massive, a smaller fraction of
their area is subtended by the 50 $\hkpc$ aperture. Thus, although the
total amount of satellites in a halo increases roughly linearly with host
halo mass, the difference in $\lsat$ between different halo masses gets
smaller. From the shaded contours, it is clear that this turnover is
statistically significant.

The right-hand panel in Figure \ref{f.lsat_propx} compares the model
predictions to the measurements of $\lsatchibar$ for star-forming and
quiescent central galaxies. For the quiescent sample, the the model is
able to fit both the steep slope in $\lsat$ for galaxies within
$1$-$\sigma$ of the mean $\cgal$, as well as the plateaus in $\lsat$ at
the high and low values of $\chi$. For the star-forming galaxies, the
models produce a monotonically increasing $\lsat$ with $\chi$, which
is in good agreement with the data at high $\chi$, but cannot fit the
data at low $\chi$, where $\lsat$ rises back up again.

Although the best-fit group catalog yields a reasonable description of
the data, the largest discrepancies are with the clustering of faint
red galaxies, the $\lsat$ values of the most concentrated quiescent
galaxies, and $\lsat$ values of the least concentrated star-forming
galaxies.

\begin{figure}
  \epsscale{1.}
  \hspace{-0.4cm}
  \plotone{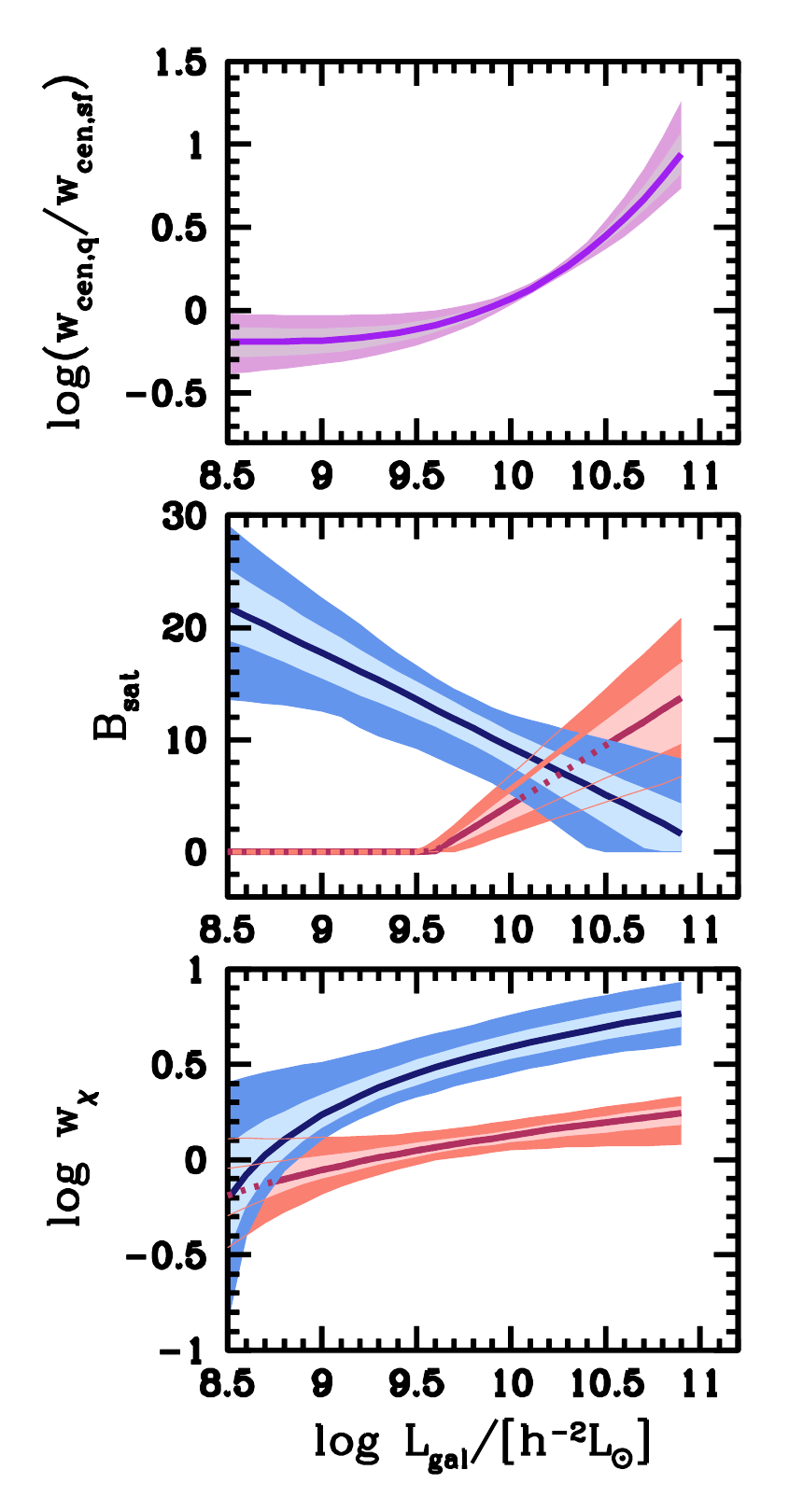}
  \vspace{-0.1cm}
  \caption{\label{f.params} Constraints on the quantities that set the
    halo masses of groups and the membership probabilities of
    satellites. See \S \ref{s.methods_params} for all equations that
    utilize the parameters. {\it Top Panel:} Constraints on the ratio
    of $\wred$ to $\wblue$. The solid curve is the best-fit group
    catalog, while the inner and outer shaded regions show the 68\%
    and 95\% value ranges. {\it Middle Panel:} Same as the top panel,
    but now for the satellite threshold value, $\bsat$. Blue and red
    contours represent results for the star-forming and quiescent
    galaxy samples, which are parameterized independently. {\it Bottom
      Panel:} Same as the middle panel, but here the contours
    represent the weights based on $\chi$, the normalized galaxy
    concentration parameter. }
\end{figure}

\subsection{Parameter constraints}

The 68\% confidence intervals on all the free parameters of the model,
as well as the values of the best-fit model, are listed in Table
1. Corner plots showing the posterior distributions of the parameters
are shown in the Appendix. 

Figure \ref{f.params} shows the constraints on the thee quantities
that determine group membership and halo mass. The top panel shows the
ratio of $\wcenr$ to $\wcenb$ for groups. Although we parameterize
these two quantities individually for maximal flexibility, the ratio
is the quantity that matters for the rank-ordering of the groups
themselves. At $\log\lcen\lesssim 10.2$, groups with quiescent central
galaxies are down-weighted relative to groups with star-forming central
galaxies. This means that, at fixed $\ltot$, quiescent-central groups
receive lower halo masses. This reverses at higher central
luminosities, where groups with star-forming centrals are
de-weighted. This yields the fit to the $\lsatred/\lsatblue$ data---at
fixed $\lcen$ (or $\ltot$), $\lsatred$ is higher because these groups
are assigned higher halo masses during rank-ordering. At lower masses,
the range of $\lsat$ ratios gets large due to weaker constraints on
the weights and few quiescent central galaxies. The best-fit model,
however, still yields $\lsatred>\lsatblue$ due to the different
scatters between the two samples, which we will discuss in \S
\ref{s.results_scatter}.

The middle panel shows the constraints on $\bsat$ for star-forming and
quiescent galaxies. Recall that the standard implementation of the
halo-based group finder set $\bsat=10$. For star-forming galaxies, $\bsat$
decreases with increasing luminosity, being nearly twice as high as
the standard value at the lowest luminosities considered. At the
bright end, $\bsat$ approaches zero.

For quiescent galaxies, the clustering requires that $\bsat$ be at the
lower limit of 0.01 for galaxies fainter than $\log\lgal=9.5$. $\bsat$ has a
linear parameterization, so for any combination of parameters that yield
$\bsat$ lower than the minimum value, $\bsat$ is simply set to the
minimum. Thus, large changes in the slope and intercept of $\bsat$
may yield no differences in the actual value used for low-luminosity
quiescent galaxies.

The bottom panel shows the constraints on $\wchi$ for star-forming and
quiescent central galaxies. The weights for quiescent centrals are
significantly lower than for star-forming centrals, implying a tighter
relationship between halo mass and $\cgal$ for quiescent
galaxies. Additionally, the $\wchi$ values for both subsamples of
galaxies show positive slopes, indicating a stronger correlation with
$\mhalo$ at lower luminosities.

\begin{figure*}
  \epsscale{1.}
  \hspace{-0.4cm}
  \plotone{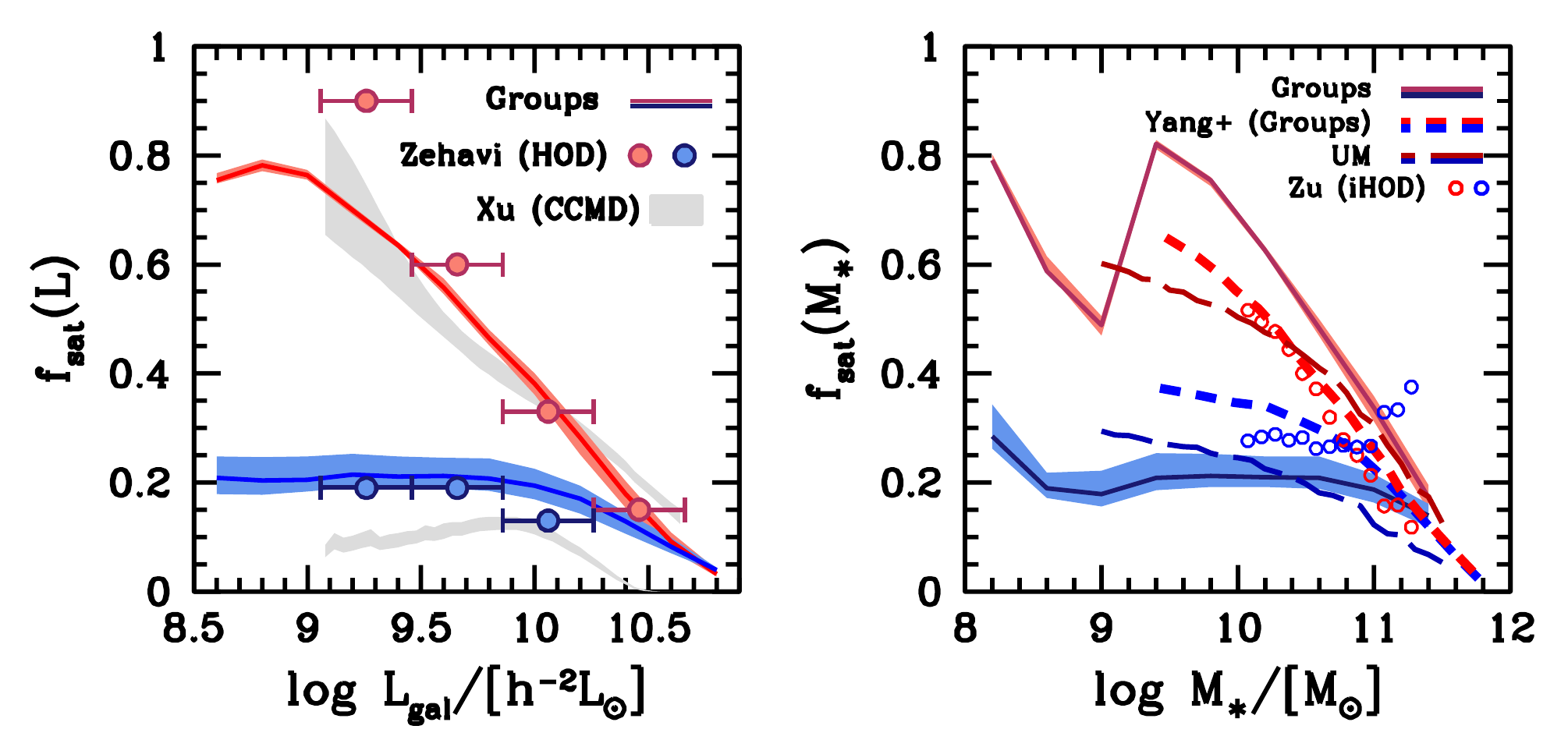}
  \vspace{-0.1cm}
  \caption{\label{f.fsat} {\it Left Panel:} The satellite fraction,
    $\fsat$, for star-forming and quiescent galaxies, as a function of
    $\lgal$. The blue and red solid curves show the results for the
    best-fit group catalog, while the shaded regions show the 95\%
    confidence interval on this quantity. The red and blue circles
    show the results of the HOD fitting of SDSS galaxies from
    \cite{zehavi_etal:11}. The gray shaded regions show the results
    for the conditional color magnitude analysis of the SDSS data by
    \cite{xu_etal:18}, which are not color-coded but are also broken
    in to star-forming and quiescent samples. {\it Right Panel:} Same
    as the opposite panel, but now for stellar mass. Results in this
    panel are $1/\vmax$-weighted (see text for details). These results
    are compared to the \cite{yang_etal:08} group catalog, the halo
    occupation analysis of \cite{zu_mandelbaum:16}, and the
    predictions of the UniverseMachine empirical model
    (\citealt{behroozi_etal:19}). }
\end{figure*}

\begin{figure*}
  \epsscale{1.}
  \hspace{-0.4cm}
  \plotone{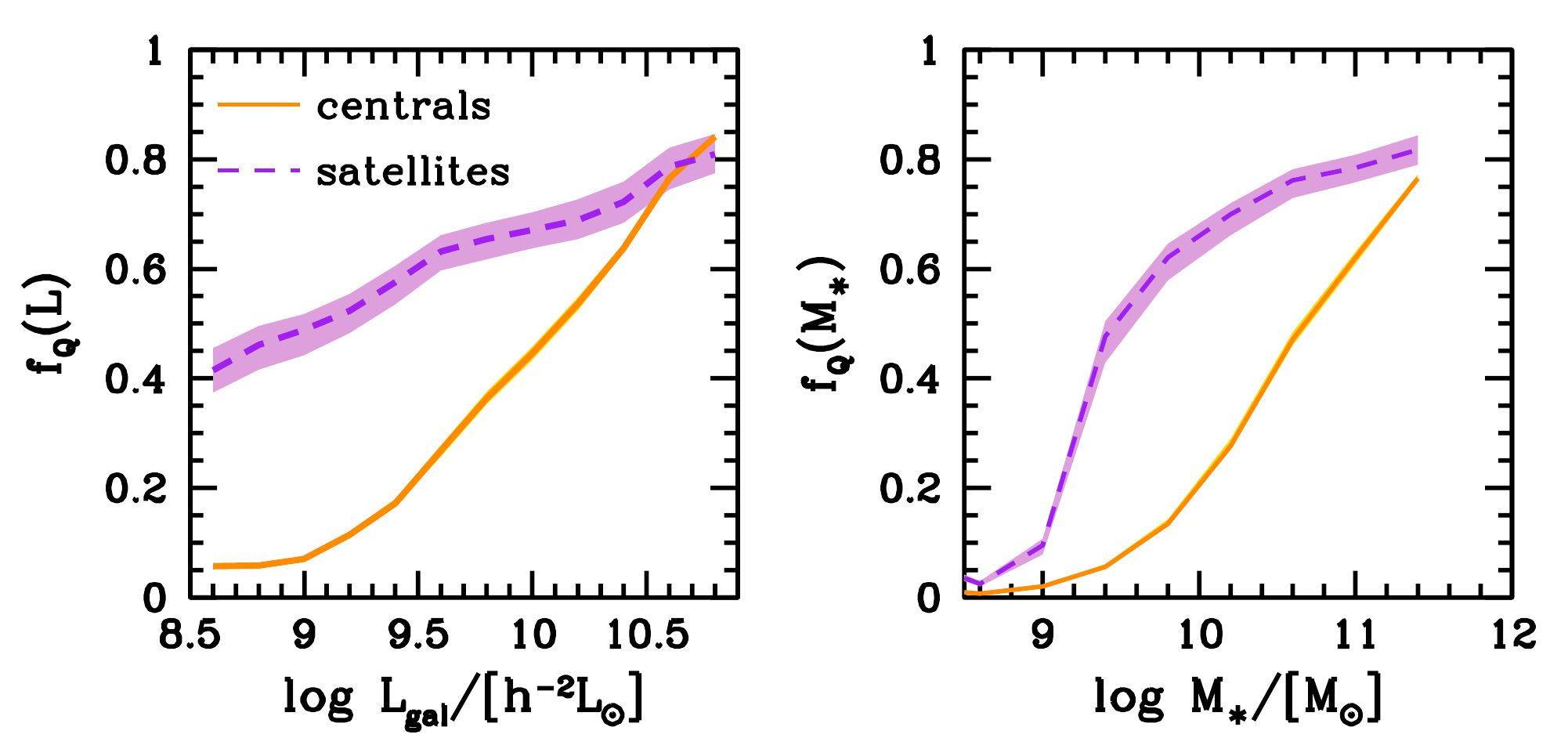}
  \vspace{-0.1cm}
  \caption{\label{f.fq} {\it Left Panel:} Quiescent fractions of
    central and satellite galaxies as a function of $\lgal$. The solid
    and dashed curves show the results for the best-fit group catalog,
    while the shaded regions show the 95\% confidence regions. {\it
      Right Panel:} Same as the opposite panel, but now as a function
    of $\mgal$. For the $\mgal$ results, all galaxies are
    $1/\vmax$-weighted. }
\end{figure*}

\begin{figure*}
  \epsscale{1.}
  \hspace{-0.4cm}
  \plotone{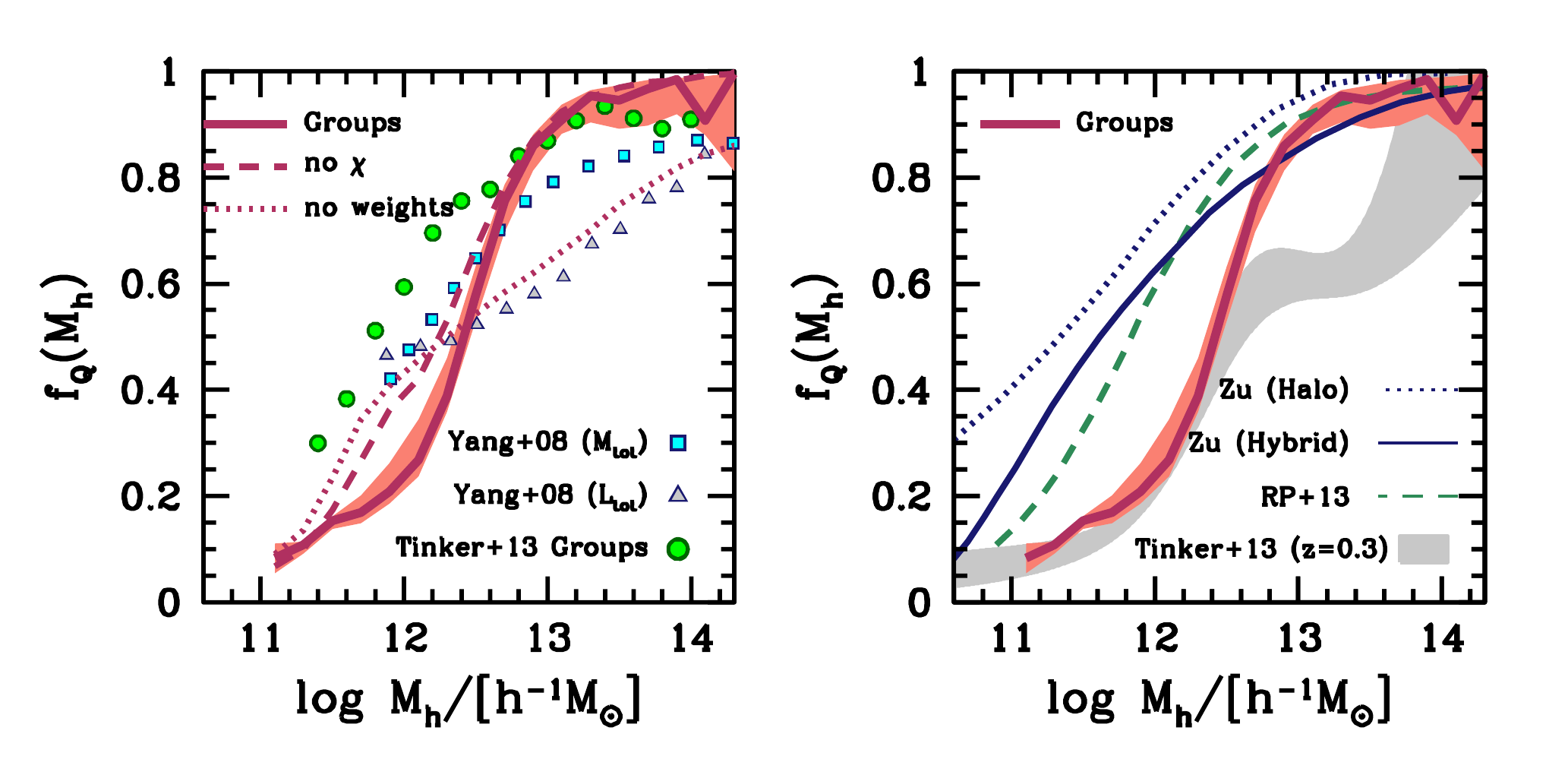}
  \vspace{-0.1cm}
  \caption{\label{f.fqh} {\it Left Panel:} Quiescent fractions of
    central and satellite galaxies as a function of $\lgal$. The solid
    curves shows the results for the best-fit group catalog, while the
    shaded regions show the 95\% confidence regions. The dashed curve
    shows the group results without $\lsatchibar$ data, and the dotted
    line is the group catalog when removing all weights from
    $\lgrp$. The symbols are previous group catalogs, described in the
    text. {\it Right Panel:} Same as the opposite panel, but now as a
    function of $\mgal$. For the $\mgal$ results, all galaxies are
    $1/\vmax$-weighted. The three curves are gray shaded region are
    halo-occupation based analysis: two parameterizations of quenching
    from \cite{zu_mandelbaum:16}, the HOD results of
    \cite{rodriguez_puebla_etal:15}, and $z=0.3$ halo occupation
    results from \cite{tinker_etal:13}.}
\end{figure*}

\subsection{Satellite fractions and quiescent fractions}

We notw turn to the constraints on the galaxy-halo connection inferred
by the self-calibrated group finder. Figure \ref{f.fsat} shows the
satellite fractions, $\fsat$ of star-forming and quiescent
galaxies. These values are presented as functions of both $\lgal$ and
$\mgal$. The shaded regions indicate the 95\% confidence intervals on
$\fsat$. The $\fsat$ constraints for quiescent galaxies are
significantly stronger than for star-forming galaxies. This is due to
the constraints on $\bsat$ presented in the previous subsection. There
is essentially no variation of $\bsat$ for quiescent galaxies fainter
than $\log\lgal=9.6$, thus the same subset of quiescent galaxies are
labeled as satellites for each model in the posterior.

We also compare our results to other analyses. For $\fsat(\lgal)$, we
compare to the HOD analysis of \cite{zehavi_etal:11}, and the
conditional color-magnitude diagram (CCMD) of \cite{xu_etal:18}. Both
are based on matching clustering and abundances of galaxies in the
MGS. The group catalog results are in good agreement with the
\cite{zehavi_etal:11} results, with the exception of the faintest bin
of quiescent galaxies, where the HOD analysis yields
$\fsat\approx0.9$. This is the magnitude bin where the group catalog
poorly fits the amplitude of the quiescent galaxy clustering. While
the prediction of the group catalog is constrained by the actual
spatial distribution of SDSS galaxies, in the HOD analysis $\fsat$ is
a completely free parameter, and thus has the freedom to increase the
satellite fraction to whatever level is required to match the
clustering. However, as is clear from the $\bsat$ constraints, there
simply are not enough quiescent galaxies in the proximity of groups to
reach such a high $\fsat$. We will discuss possible improvements to
the self-calibrated group finder in \S \ref{s.discussion}.

Our $\fsat(\lgal)$ results are in good agreement with the CCMD results
for quiescent galaxies, but the CCMD satellite fraction of star-forming
galaxies is lower than both the group catalog and the HOD results by
nearly a factor of two. We will discuss this further in section \S
\ref{s.results_scatter}, and how it relates to constraints on scatter
in the galaxy-halo connection.

The right-hand side of Figure \ref{f.fsat} shows the group catalog
results for $\fsat(\mgal)$. As opposed to the results for $\lgal$,
where each bin in luminosity is effectively a volume-limited sample,
each bin in $\mgal$ contains galaxies of varying luminosity, and thus
various $\vmax$ values. To account for this, in each bin
of $\mgal$, $\fsat$ is
calculated as 

\begin{equation}
  \label{e.fsat_vmax}
  \fsat(\mgal) = \frac{\sum_{\rm sats} 1/\vmax}{\sum_{\rm all}
    1/\vmax}
\end{equation}

\noindent where the sums in the numerator and the denominator are over
all satellites and all galaxies within the stellar mass bin. As with
the $\fsat$ results for luminosity, the constraints for quiescent
galaxies are significantly tighter than those for star-forming
galaxies. The abrupt change in quiescent $\fsat$ at $\mgal<10^9$
$\msol$ is due to finite number statistics: although all results are
$1/\vmax$ weighted, the raw number of quiescent galaxies below this
stellar mass is only 146.

This figure compares the self-calibrated group catalog results to
those of the \cite{yang_etal:08_hod}. As demonstrated in Paper I, the
use of $\bsat=10$ for all galaxies yields significant biases in
$\fsat$ which are mirrored in this result: $\fsat$ for quiescent
galaxies is suppressed while the number of star-forming satellites is
increased. We also compare to the results of the iHOD analysis of
\cite{zu_mandelbaum:16}. The iHOD analysis is analytic, but the
$\fsat$ results, kindly provided by Y.~Zu, are measured from a
populated N-body simulation. Thus the results for bright blue galaxies
are subject to high shot noise, but are otherwise in agreement with
the results from the Yang catalog.

We also compare our results to predictions of the
UniverseMachine empirical model of \cite{behroozi_etal:19}, taken from
the {\sc mockUM} prepared for testing in Paper I. The satellite fraction for
star-forming galaxies is in reasonable agreement with the
self-calibrated group results, but $\fsat$ for quiescent galaxies is
somewhat lower.

Figure \ref{f.fq} shows a complementary statistic: the {\it quiescent
  fraction}, $\fq$, of central and satellite galaxies. As quenching efficiency
is higher for satellite galaxies, the expectation is that a higher
fraction of satellites will be quiescent than central galaxies (e.g.,
\citealt{weinmann_etal:06, wetzel_etal:12, wetzel_etal:13,
  behroozi_etal:19}). The group catalog yields enhanced $\fq$ values
for satellites, when binned both by $\lgal$ and by $\mgal$. For
$\fq(\lgal)$, the luminosity at which half of centrals are quiescent
is $\log\lgal=10.1$, while for satellites this value is
$\log\lgal=9.2$. For $\fq(\mgal)$, these values for centrals and
satellites are $\log\mgal=10.6$ and 9.4, respectively.

In contrast to the $\fsat$ results, $\fq$ for satellites has larger
uncertainties than $\fq$ for centrals, even though the 95\%
constraints for both are less than a few percent. But $\fq$ for
satellites is determined by the uncertainties in all the probabilities
for galaxies to be labeled satellites, and $\bsat$ for star-forming
galaxies has significantly more uncertainty than for quiescent
galaxies. For central galaxies, the 95\% confidence intervals are
barely wider than the curve showing the best-fit model. Although there
is some uncertainty on what gets labeled a satellite galaxy,
satellites themselves make up $<1/3$ of the overall galaxy population,
thus the constraints on $\fq$ for centrals will be stronger by
definition.

\subsection{Quenching as a function of halo mass}

Rather than bin galaxies by their observed properties, it is of
interest to investigate how the quiescent fraction of central galaxies
depends on {\it halo mass}. This is one of the main quantities for
which previous group catalogs have shown significant biases in
\cite{campbell_etal:15}, and one of the goals of the
self-calibrated algorithm.

Figure \ref{f.fqh} shows $\fq$ for central galaxies as a function
of $\mhalo$. The comparison is split into two panels---in each panel
the fiducial group catalog results are the same, but in the left-hand
panel we compare to previous group results while on the right we
compare our new group results to those constrained by fitting halo
occupation models to measurements of clustering and lensing.

First, the left-hand panel of Figure \ref{f.fqh} shows how our group
results depend on different assumptions in the implementation of the
group finder. The thick solid line shows the best-fit group catalog
when considering all data. The dashed curve shows the no-$\chi$ group catalog
results, meaning
all $\wchi$ weights are not included in the rank-ordering of the
groups. Removing this freedom from the catalog has negligible impact
on the results at $\mhalo\gtrsim 10^{12.5}$ $\hmsol$, but it changes
the slope of $\fq$ at lower masses, making the transition from
mostly-quiescent central galaxies to mostly-star-forming centrals more
gradual. The dotted line is removing all weights of any kind, as well
as setting $\bsat=10$ for all galaxies. This is effectively the same
as the standard halo-base group finder, if the rank-ordering of the
groups is based on group luminosity.

The comparison with the gray triangles confirms this, as these data
are from the \cite{yang_etal:08} group catalog where group ranking is
based on $\ltot$. The blue squares are the same catalog, but now
ranking my total stellar mass of the group instead. As quiescent
galaxies at fixed stellar mass have lower luminosities, the blue
squares have a much higher $\fq$ at high halo masses, getting closer
to the values found in our self-calibrated group finder that excluded
$\lsat(\chi)$ (dashed line). Last, the green circles show the results
of our previous halo-based group finder, with results presented in
\cite{tinker_etal:13}. These results differ from the others in
that they are performed on volume-limited samples, complete in stellar
mass. Here, $\fq$ is in good agreement with the self-calibrated groups
at high halo masses, but once again has a softer transition to groups
with majority star-forming central galaxies.

The right-hand panel in Figure \ref{f.fqh} compares our group catalog
to HOD-based methods. The \cite{zu_mandelbaum:16} and
\cite{rodriguez_puebla_etal:15} analysis are both performed on the
SDSS MGS, while the \cite{tinker_etal:13} results are from
analysis of clustering and lensing within the COSMOS field.  
\cite{zu_mandelbaum:16} present two analyses; one in which quenching
is entirely a halo mass-driven process, and another in which quenching
is a `hybrid' combination of halo and stellar
masses. \cite{rodriguez_puebla_etal:15} also parameterize $\fq$ as a
function of $\mhalo$. All three of these models yield quiescent
fractions of halos at $\mhalo\lesssim 10^{12.5}$ $\hmsol$
significantly higher than that found in the group catalog.

\cite{tinker_etal:13} implement a non-parametric model for
$\fq(\mhalo)$ that allows for the non-monotonic behavior seen in the
constraints in Figure \ref{f.fqh}. At the highest mass scales and at
$\mhalo\lesssim 10^{12.5}$ $\hmsol$, these results are in good
agreement with the self-calibrated groups. But the small area of the
COSMOS field---which contributes to the unusual behavior of the
$\fq(\mhalo)$ constraints is an extra source of uncertainty.

\begin{figure*}
  \epsscale{1.}
  \hspace{-0.4cm}
  \plotone{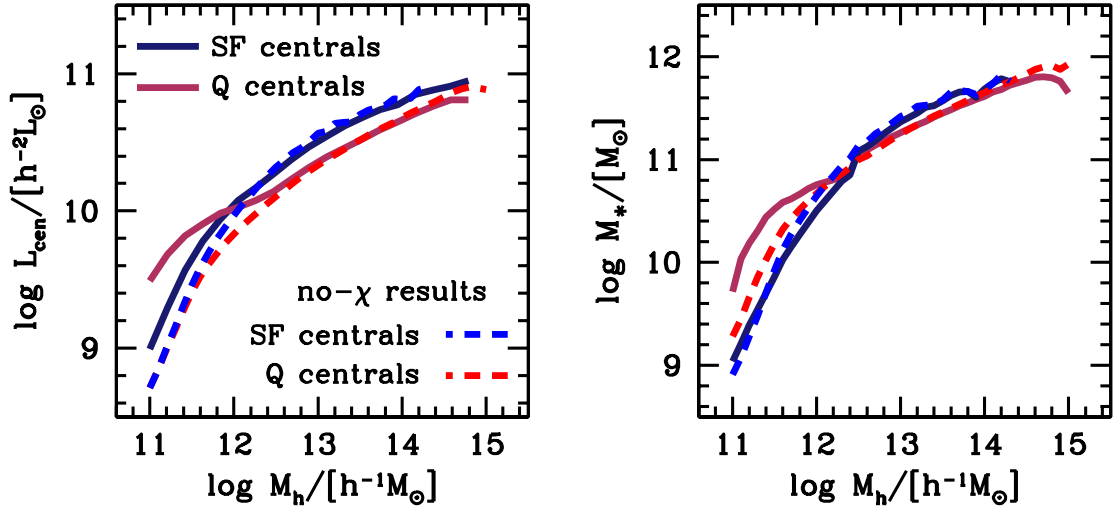}
  \vspace{-0.1cm}
  \caption{\label{f.shmr2} {\it Left Panel:} LHMR for star-forming and
    quiescent central galaxies from the group catalog. We do not show
    uncertainties for the quantity because they are nearly the width
    of the surves for most halo masses. The dashed lines show the
    results from the group catalog that does not include the
    $\lsat(\chi)$ data. Red colors indicate results for quiescent
    galaxies, while blue colors show results for star-forming
    galaxies. {\it Right Panel:} Same as the opposite panel, but now
    replacing galaxy stellar mass, $\mgal$, for luminosity. }
\end{figure*}

\subsection{LHMRs and SHMRs}
\label{s.results_lhmrs}

Figure \ref{f.shmr2} shows the LHMR and the SHMR predicted by the
self-calibrated group finder. We will compare these results to
previous results in subsequent figures. For the LHMR, there is a clear
change at $\mhalo \sim 10^{12}$ $\hmsol$: not only is this is location
of the `pivot scale,' where $\lgal/\mhalo$ is maximal, but the LHMRs
for star-forming and quiescent galaxies cross. Above this halo mass
scale, central star-forming galaxies are more massive than quiescent
centrals at the same $\mhalo$. Below this scale, this trend reverses
and quiescent galaxies are more massive at fixed $\mgal$. When binned
by $\lcen$, as in the comparison to the $\lsatred/\lsatblue$ data in
Figure \ref{f.lsat_propx}, the {\it halos} of quiescent galaxies are
more massive than star-forming galaxies at all $\lcen$. As we will see
in the following subsection, the scatter of $\log\lgal$ at fixed
$\mhalo$ changes significantly across the halo mass spectrum. The
larger scatter at low $\mhalo$ makes a significant difference in the
statistics, depending on whether they are binned by $\mhalo$ or by
$\lcen$.

The impact of scatter on the LHMR is also apparent when comparing the
best-fit group catalog to our test catalog where we do not include the
$\lsat(\chi)$ measurements. As discussed in Paper I, without these
data, the group catalog cannot differentiate the halos of galaxies at
low luminosities. Thus the inferred scatter approaches zero. The
dashed lines in Figure \ref{f.shmr2} show the LHMRs for this
catalog. Above the pivot point, there is little difference in the
results. At low $\mhalo$, however, the LHMRs for star-forming and
quiescent samples come together, and both are offset from the fiducial
catalog.

The right-hand side of Figure \ref{f.shmr2} shows the same results,
but for $\mgal$ as opposed to $\lgal$. Above the pivot mass, the SHMRs
for star-forming and quiescent centrals are closer together---as
expected from the fact that quiescent galaxies are fainter at fixed
$\mgal$---but the SHMR for star-forming galaxies is still higher than
that of quiescent centrals. The switch below the pivot mass is even
more pronounced---at the lowest halo masses probed, quiescent central
galaxies are nearly a dex more massive than star-forming central
galaxies at the same $\mhalo$.

\begin{figure}
  \epsscale{1.2}
  \hspace{-0.4cm}
  \plotone{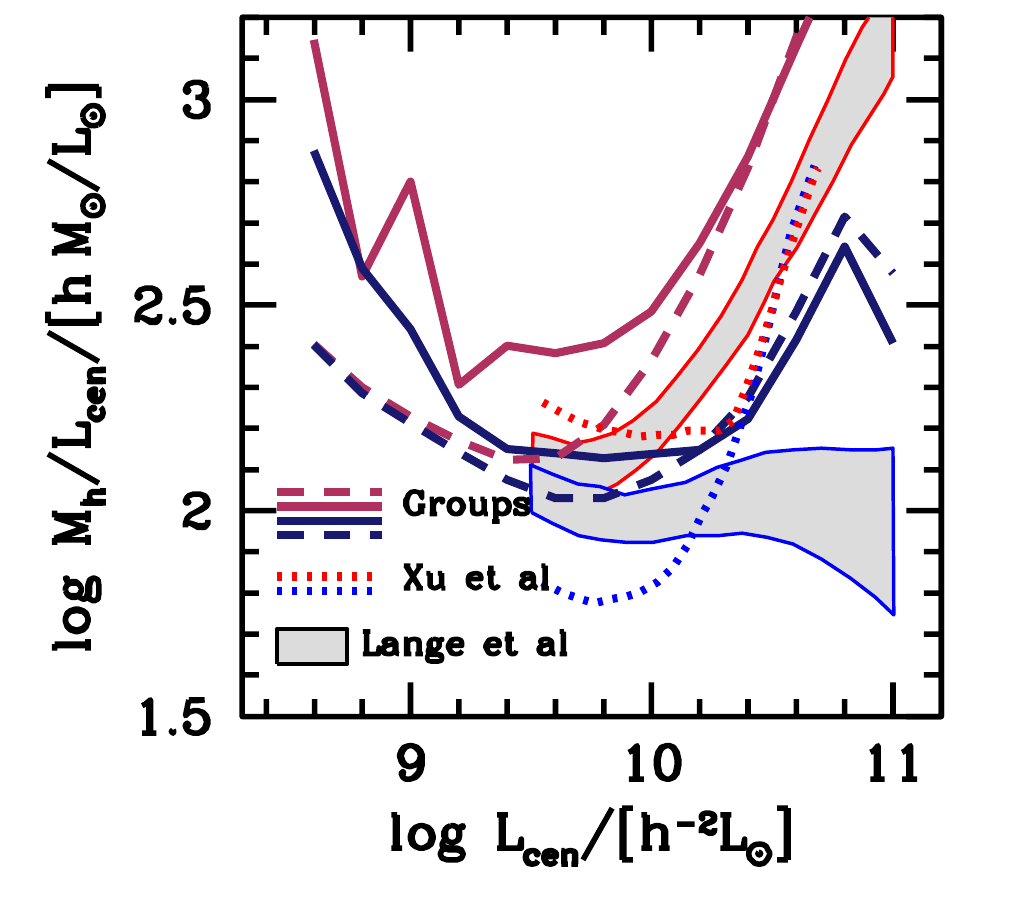}
  \vspace{-0.8cm}
  \caption{\label{f.ml_ratio} Comparing the halo mass-to-light ratio
    of central galaxies, $\ml$, of the self-calibrated group catalog
    to other results. The thick solid curves show the fiducial group
    catalog, where star-forming and quiescent centrals are indicated
    by the dark blue and dark red colors. The dashed curves with the
    same colors indicate the group catalog results that do no include
    the $\lsat(\chi)$ data. The shaded region shows the results of the
    satellite kinematics analysis of \cite{lange_etal:19}. The range
    indicates the 95\% confidence interval from their analysis. The
    border color corresponds to red and blue galaxy subsamples. The
    dotted curves show the conditional color-magnitude diagram results
    of \cite{xu_etal:18}. We have corrected the halo masses of two
    previous results to account for the difference between
    $M_{\rm vir}$ and $M_{200b}$.}
\end{figure}

\subsection{Mass-to-Light ratios}
\label{s.results_ml}

Figure \ref{f.ml_ratio} shows the mass-to-light ratios for
central galaxies, $\ml$, broken in to our star-forming and quiescent samples,
plotted as a function of $\lcen$. We show both the results from our
best-fit group catalog as well as the no-$\chi$ catalog. The results
of the two catalogs diverge at $\lcen<10^{10}$ $\lsolhh$, where the
fiducial catalog yields higher $\ml$ ratios. We also note that, at low
luminosities, the results of the fiducial catalog show only a small
difference in the M/L ratios of star-forming and quiescent samples,
analogous to the $\lsat$ results in Figure \ref{f.lsat_propx}
discussed earlier.

We compare these results to those of \cite{lange_etal:19} and the CCMD
results of \cite{xu_etal:18}. \cite{lange_etal:19} use a halo
occupation model to fit measurements of satellite kinematics
data. Central galaxies are binned by luminosity and separated into red
and blue subsamples using the \cite{zehavi_etal:11} tilted color cut
described in \S \ref{s.data_sfq}. The \cite{lange_etal:19} results
show the same bimodality in halo masses between star-forming and
quiescent samples as seen in the group catalog, although the overall
amplitude is shifted down relative to the groups by $\sim 0.2$
dex. The \cite{xu_etal:18} results, also separated using the tilted
color cut, show color-dependent bimodality in halo
masses  at $\lcen\lesssim 10^{10}$ $\lsolhh$. At higher
luminosities, the $\ml$ ratios converge. But the consensus from these
results---all based on MGS data, binned by $\lgal$ and separated into
star-forming and quiescent samples, but using complementary methods to
connect galaxies to halos---is that the $\ml$ for quiescent galaxies
is equal to or above that of star-forming centrals.

\begin{figure*}
  \epsscale{1.0}
  \hspace{-1.4cm}
  \plotone{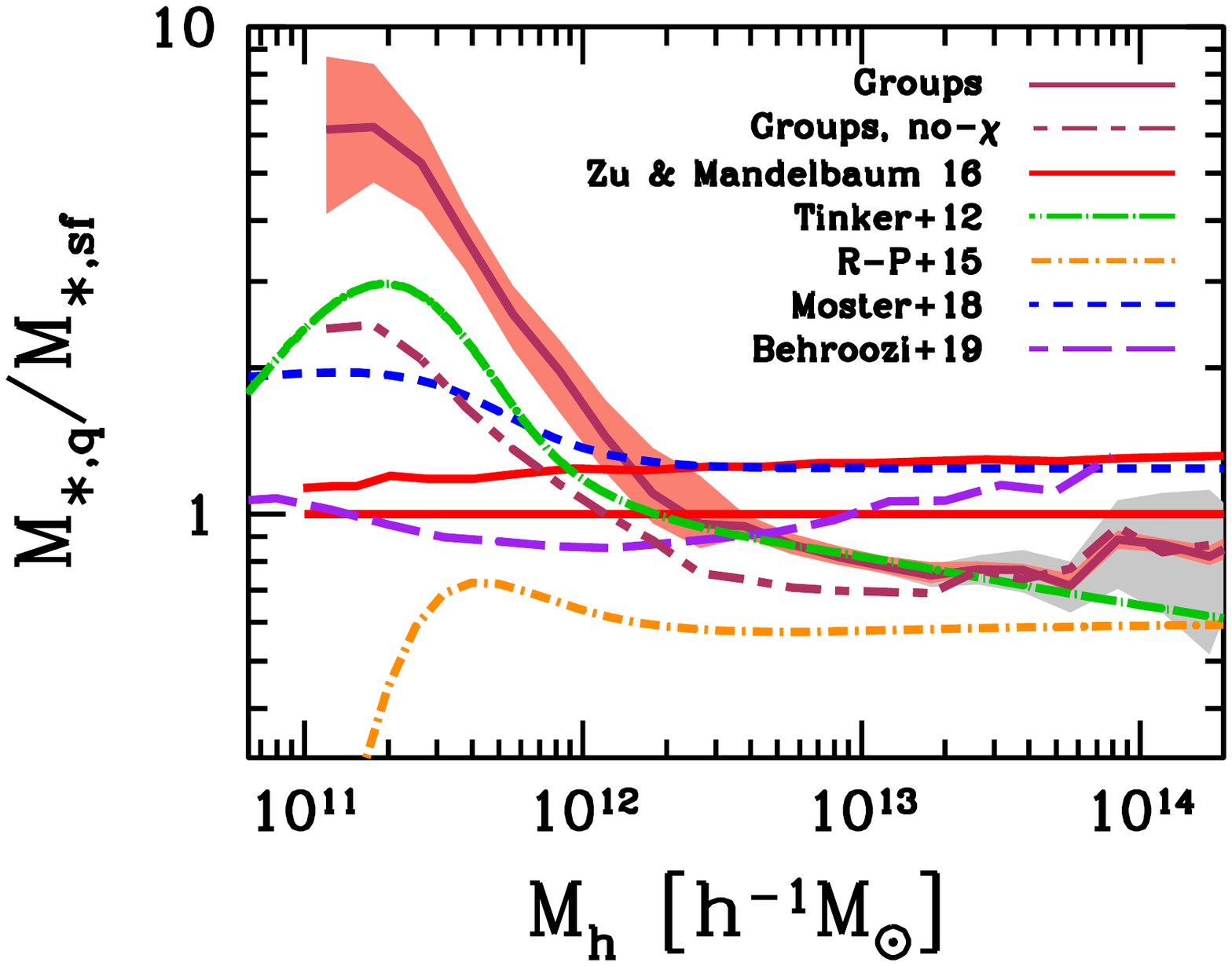}
  \vspace{-0.2cm}
  \caption{\label{f.redblue_ratio} The ratio of stellar masses for
    quiescent and star-forming central galaxies, $\mqmsf$, as a
    function of $\mhalo$. The solid dark red curve shows the fiducial
    group catalog results. The red shaded region shows the 95\%
    confidence interval from the MCMC chain, while the gray shaded
    region shows the $2\sigma$ Poisson error bars, which only become
    significant at large halo masses. The dashed curve of the same
    color shows the group results excluding the $\lsat(\chi)$
    data. The other curves show constraints on this quantity from halo
    occupation and empirical models. }
\end{figure*}

\begin{figure}
  \epsscale{1.3}
  \hspace{-1.2cm}
  \plotone{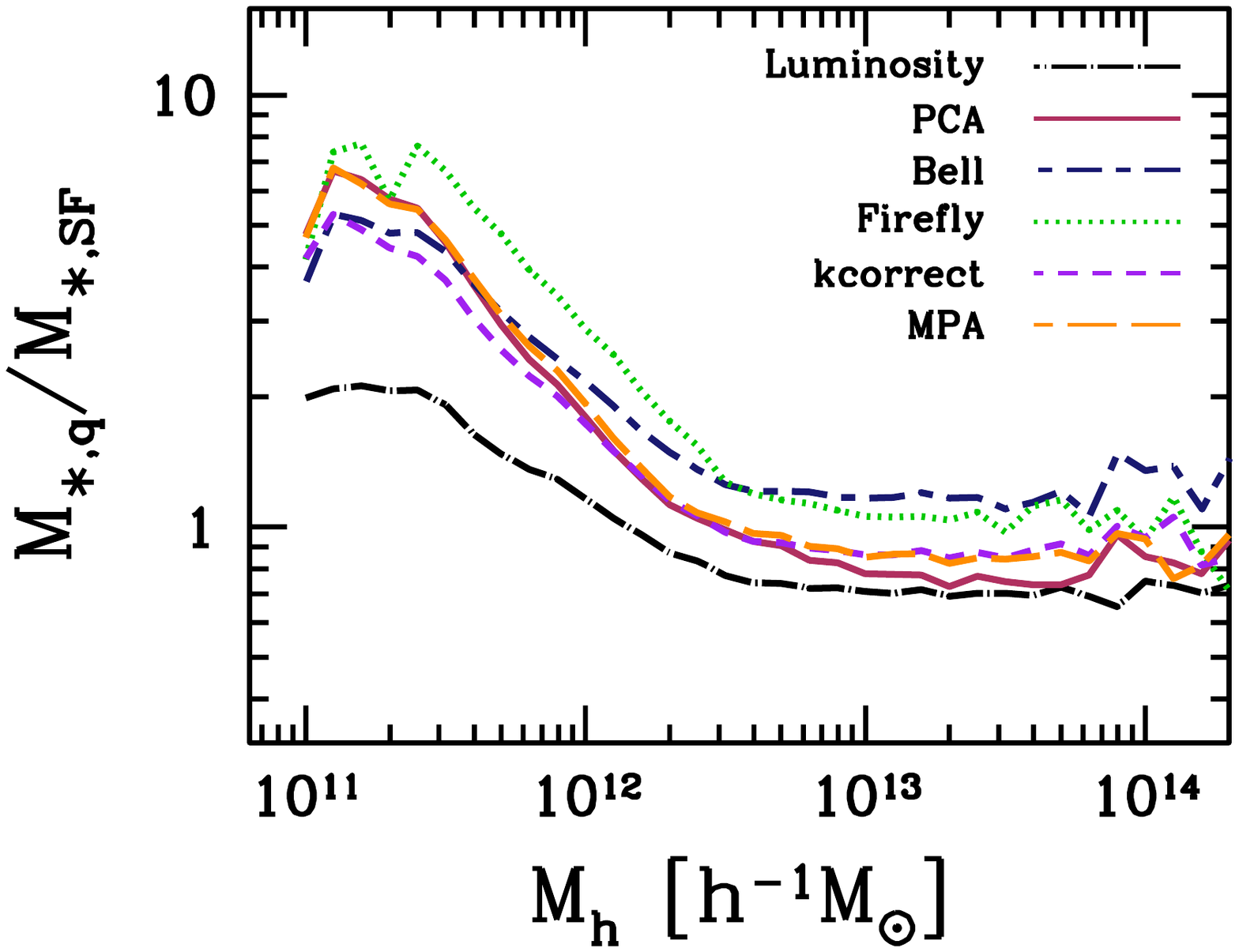}
  \vspace{-0.2cm}
  \caption{\label{f.redblue_masses} The results of Figure
    \ref{f.redblue_ratio}, but now changing the stellar masses used in
    the calculation. The PCA results are our fiducial results. These
    masses and those from the Firefly code use the full spectra of the
    galaxy to estimate $\mgal$. The {\tt kcorrect} masses use only the
    SDSS broadband magnitudes. The MPA masses use a combination of
    broadband imaging and spectra features. Each show nearly the same
    trend with $\mhalo$, but with somewhat varying amplitudes. For
    reference, the ratio of $\lcen$ for quiescent and star-forming
    galaxies is also shown.}
\end{figure}

\subsection{Stellar masses of central galaxies}

Figure \ref{f.redblue_ratio}, adapted from Figure 8 in
\cite{wechsler_tinker:18}, shows the stellar mass ratio of
star-forming and quiescent central galaxies, $\mqmsf$,  as a function of
$\mhalo$. We show both the fiducial and the no-$\chi$ group
results. For both of these catalogs, the $\mqmsf$ ratio is below unity
at high halo masses and above unity at low halo masses. The two
catalog agree at high halo masses, but at $\mhalo\lesssim 10^{13}$
$\hmsol$ the fiducial results are higher than the no-$\chi$ results.

Figure \ref{f.redblue_ratio} shows the constraints on this quantity
from various analyses in the field. The results range from $\mqmsf$
being below unity (\citealt{rodriguez_puebla_etal:15}), above unity
(\citealt{moster_etal:18}), and consistent with unity
(\citealt{zu_mandelbaum:16, behroozi_etal:19}). The COSMOS analysis of
\cite{tinker_etal:13} is in qualitative agreement with the groups
results, in that the mass ratio switches from $>1$ to $<1$ at roughly
the same halo mass scale. The main reason why this is notable is
that it is possible to get such a sign change from a halo occupation
analysis, which is not clearly exhibited in the other results.

We further note that the increase in $\mqmsf$ at low halo masses is, in
part, driven by the choice of $\ltot$ to rank-order groups as opposed
to $M_{\ast,tot}$. As seen in
Figure \ref{f.shmr2}, for the no-$\chi$ results the LHMRs converge at
low masses. Thus, the {\it luminosity} ratio of star-forming and
quiescent central galaxies is unity at $\mhalo<10^{12}$ $\hmsol$. The
different stellar populations then yield higher stellar masses for
quiescent galaxies at fixed $\lgal$. However, Figure \ref{f.shmr2}
demonstrates that the fiducial catalog puts higher luminosity
quiescent centrals in low-mass halos. Thus, if we use $M_{\ast,\rm
  tot}$ instead of $\ltot$, the mass ratio would still be greater than
unity at low $\mhalo$, but possibly with a smaller amplitude.

Figure \ref{f.redblue_masses} compares the $\mqmsf$ ratio for the
fiducial catalog for
different stellar mass estimates available for MGS data. For
reference, this figure also shows the ratio of luminosities for
central galaxies. Here we include our fiducial PCA stellar masses, as
well as masses from the MPA-JHU catalog
(\citealt{brinchmann_etal:04}), Firefly stellar masses
(\citealt{firefly}), the stellar masses produced from the
{\tt kcorrect} code (\citealt{blanton_roweis:07}), and the M/L-ratio
corrected stellar masses of \cite{bell_etal:03}. Compared to the
luminosity ratio, all stellar mass ratios show a larger change from
low to high halo masses. The specific star formation rate of galaxies
increases with decreasing $\mgal$ (e.g.,
\citealt{noeske_etal:07b}). Thus, for star-forming galaxies at low
luminosities, a larger fraction of the luminosity comes from young
stars, even in the $r$-band. For luminous galaxies, the stellar masses
of quiescent and star-forming galaxies are closer together because
star-forming galaxies have lower specific star-formation rates. The
takeaway from Figure \ref{f.redblue_masses}, however, is that the
shape of $\mqmsf$ as a function of $\mhalo$ is roughly independent of
stellar mass method used. 

\begin{figure}
  \epsscale{1.25}
  \hspace{-1.5cm}
  \plotone{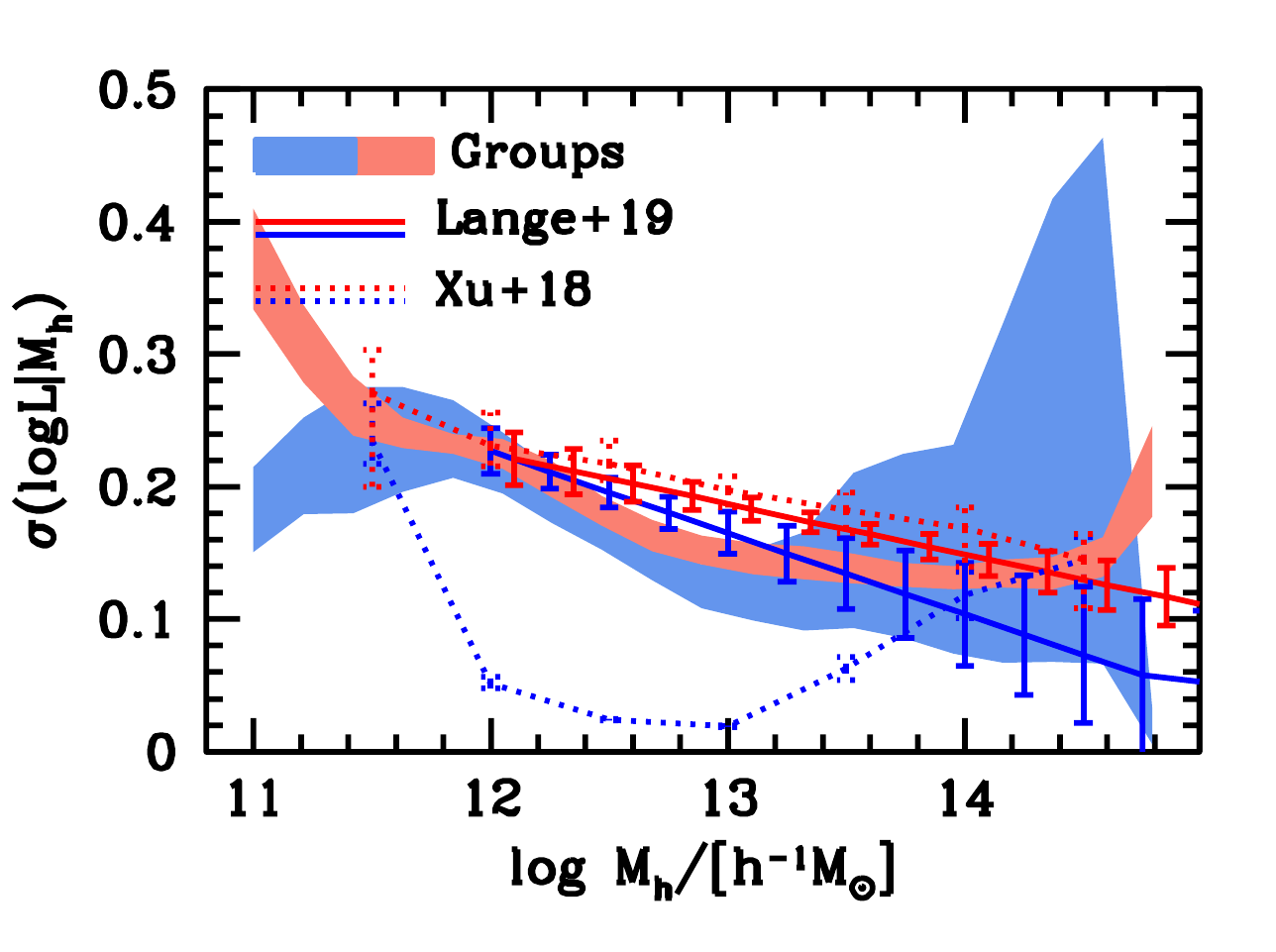}
  \vspace{-0.1cm}
  \caption{\label{f.slogl} Scatter of $\log\lcen$ as a function of
    $\mhalo$, $\slogl$. The blue and red shaded contours show the 95\%
    confidence intervals from the group catalog for star-forming and
    quiescent galaxy samples, respectively. The solid lines are the
    results from the satellite kinematics analysis of
    \cite{lange_etal:19}, in which $\slogl$ is parameterized as a
    linear function of $\log\mhalo$. Error bars represent 95\%
    confidence intervals on their results. The dotted curves are the
    results of the CCMD analysis of \cite{xu_etal:18}, which are
    driven by clustering measurements. }
\end{figure}

\begin{figure}
  \epsscale{1.25}
  \hspace{-1.5cm}
  \plotone{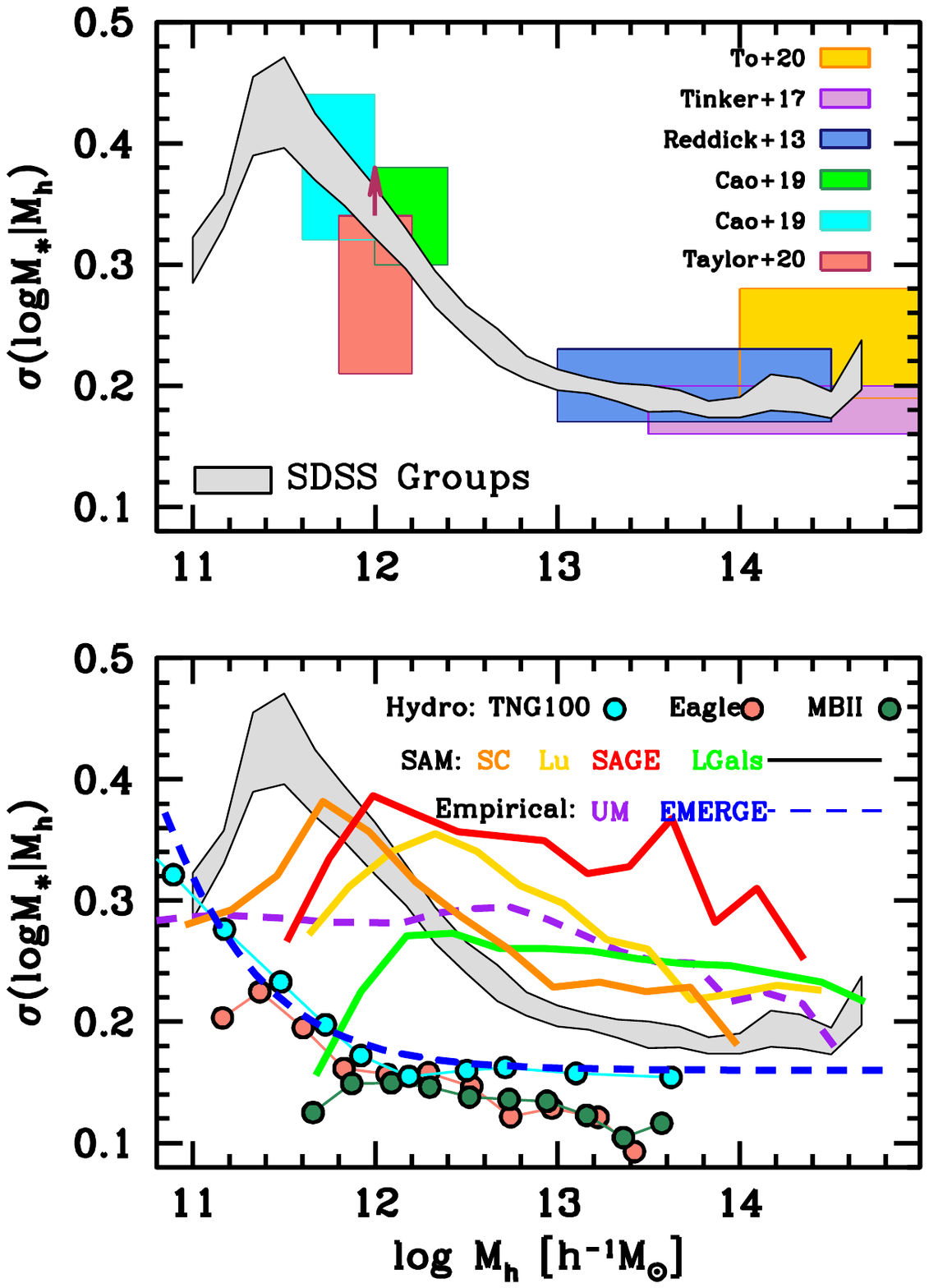}
  \vspace{-0.1cm}
  \caption{\label{f.slogm} {\it Upper Panel:} A comparison of the
    scatter in the SHMR between the group catalog and other
    observational constraints. The gray shaded region shows the 95\%
    confidence interval on $\slogm$. The observational constraints com
    from a range of sources. At high halo masses, the results come
    from clusters (\citealt{to_etal:20}), clustering
    (\citealt{tinker_etal:17_boss}), and galaxy groups
    (\citealt{reddick_etal:13}). At lower masses, two independent
    constraints from from \cite{cao_etal:19}, while the results from
    \cite{taylor_etal:20} are derived from weak lensing, and represent
    a lower limit to $\slogm$. {\it Lower Panel:} A comparison between
    group catalog and predictions from numerical simulations.  The
    filled, connected circles show results from hydrodynamic
    simulations (\citealt{tng, matthee_etal:17, khandai_etal:15}),
    semi-analytic models (\citealt{somerville_etal:08, lu_etal:14,
      henriques_etal:17, sage}, and empirical models
    (\citealt{behroozi_etal:19, moster_etal:18}).}
\end{figure}

\subsection{Scatter}
\label{s.results_scatter}

Figure \ref{f.slogl} shows the scatter in $\log\lcen$ as a function of
$\mhalo$ for the fiducial group catalog, which we refer to as $\slogl$. This
quantity is separated into star-forming
and quiescent central galaxies. For the best-fit catalog, $\slogl$
decreases with increasing $\mhalo$ for both samples. The uncertainty
in the scatter for quiescent centrals is significantly smaller than
that of star-forming centrals. For massive halos this is mostly driven
by the small numbers of star-forming central galaxies, but it is also
driven by the tighter constraints $\wchi$ for the quiescent
sample---as shown in Paper I, the value of $\slogl$ derived from the
group catalog is dependent on the correlation coefficient between
$\mhalo$ and $\chi$. Because of this, the results in Figure
\ref{f.slogl} should be considered lower limits on the scatter. Due to
the differences in $\wchi$, however, the scatter for quiescent
galaxies is likely to be much closer to the true value than for
star-forming centrals.

In this Figure we also compare to the satellite kinematics constraints
of \cite{lange_etal:19} and the CCMD constraints of
\cite{xu_etal:18}. \cite{lange_etal:19} parameterize $\slogl$ as a
linear function of $\log\mhalo$. For the CCMD, the results are more
complicated, in that \cite{xu_etal:18} parameterize the CCMD with
`pseudo'-blue and -red populations that can overlap, but are then
translated into observed color space. Scatter in the pseudo
populations are parameterized in the same log-linear fashion as
\cite{lange_etal:19}. In general, all three methods yield consistent
constraints, with scatter values centered on $\slogl\sim 0.2$ dex, and
with decreasing scatter as a function of $\mhalo$. The exception is
the scatter in luminosity for blue galaxies in the CCMD. As explained
in Xu et.~al., the smallar scatter is required to match the clustering
amplitude of the blue galaxies. This may be related to the lack of
blue satellites in the CCMD, seen in Figure \ref{f.fsat}, which also
increase the large-scale bias of the galaxy sample.

Figure \ref{f.slogm} shows the scatter in $\log\mgal$ as a function of
$\mhalo$ for the group catalog, which we refer to as $\slogm$. Here we
show the overall scatter in order to compare with other observational
results and theoretical predictions. At large halo masses, $\slogm$
plateaus at $\sim 0.2$ dex, as compared to 0.15 dex for $\slogl$. This
reflects the scatter in $\mgal$ at fixed $\lgal$.  As with the
luminosity results, $\slogm$ increases with decreasing
$\mhalo$. However, the rise at $\mhalo<10^{13}$ $\hmsol$ is much
steeper. This is partly driven by the increased scatter between
luminosity and stellar mass for star-forming galaxies, but the
divergence at $\mhalo<10^{11.5}$ $\hmsol$ is possibly affected by
incompleteness in the flux-limited catalog. We note that the $\mgal$
results incorporate $1/\vmax$ weighting in all calculations to mimize
this effect.

In the upper panel of Figure \ref{f.slogm} we compare our results to
other observational constraints on $\slogm$. At $\mhalo>10^{13}$
$\hmsol$, we compare to estimates from galaxy clusters
(\citealt{to_etal:20}), galaxy groups (\citealt{reddick_etal:13}), and
the clustering of massive galaxies
(\citealt{tinker_etal:17_boss}). These results, as well as other
results from the literature (\citealt{yang_etal:09, leauthaud_etal:12_shmr,
  rodriguez_puebla_etal:15, zu_mandelbaum:16}) converge on a value of
$\slogm\sim 0.2$ dex for massive halos.

There are fewer observational constraints at lower halo
masses. \cite{cao_etal:19} used both clustering cross-correlations and
line-of-sight velocity distributions to produce two independent
constraints that isolate low-mass halos. Additionally,
\cite{taylor_etal:20} used galaxy-galaxy lensing to constrain a lower
limit on the scatter in the SHMR. All three of these measurements
yield higher values than that seen at high mass scales. The results
from the group catalog are in good agreement with all of these
observational constraints.

The lower panel in Figure \ref{f.slogm}, also adapted from
\cite{wechsler_tinker:18}, compares the group results to theoretical
predictions. We separate the theoretical predictions into three
classes: cosmological hydrodynamical simulations, semi-analytic model
of galaxy formation, and empirical models. There is a clear separation
between the hydrodynamic and semi-analytic predictions, with
semi-analytic models yielding higher scatter values. Both classes of
models exhibit an increase\footnote{The downturns in the predictions
  for both classes of models in the lowest one or two data points is
  likely a result of resolution limits.} in $\slogm$ at low
$\mhalo$. The hydrodynamic models show an asymptote at high masses and
a steep rise at low masses, while the semi-analytic predictions
exhibit a more linear change in scatter. The two empirical model show
disparate results, with the EMERGE results consistent with the
hydrodynamic simulations and UniverseMachine comparable to the
semi-analytic results.

\begin{figure}
  \epsscale{1.25}
  \hspace{-1.5cm}
  \plotone{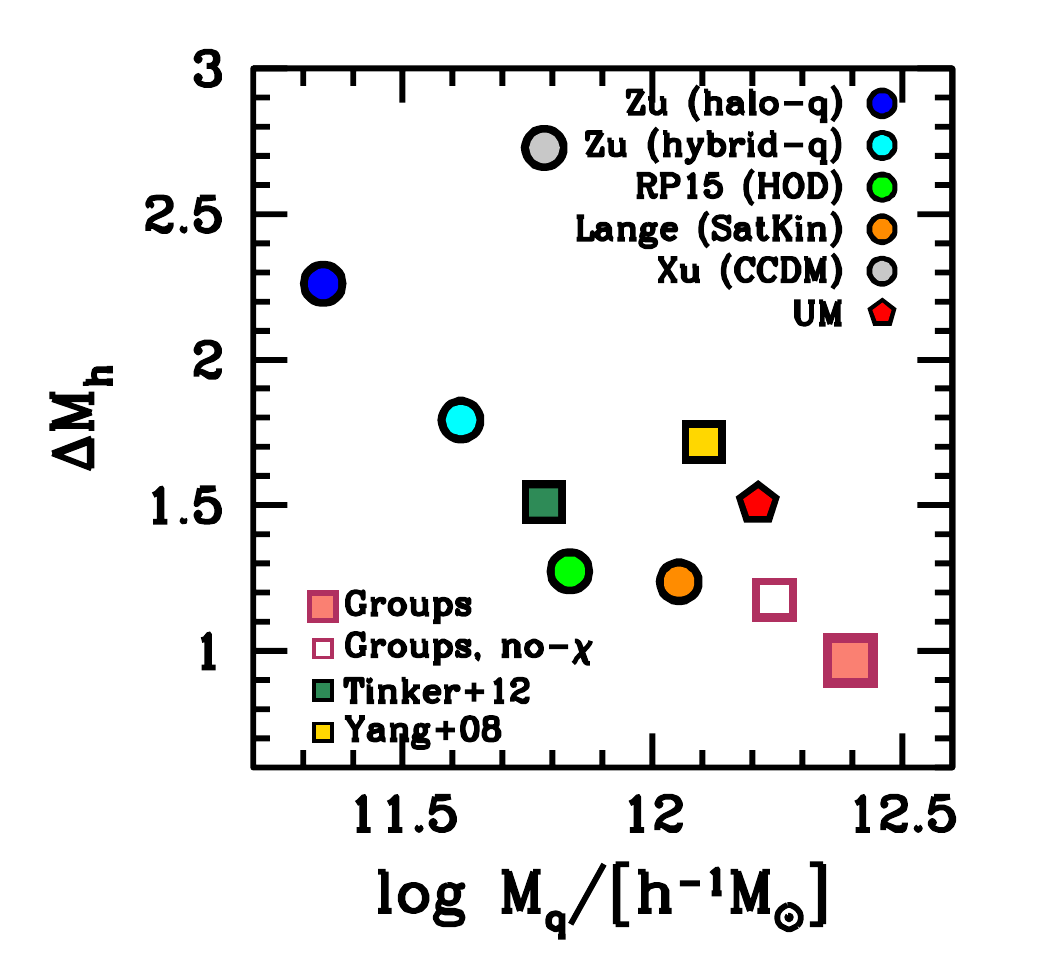}
  \vspace{-0.1cm}
  \caption{\label{f.fqh_transition} Model predictions for the
    transition between halos that contain star-forming central
    galaxies to halos with quiescent central galaxies. The $x$-axis is
    the halo mass scale at which 50\% of central galaxies are
    quiescent. The $y$-axis is the range in $\log\mhalo$ over which
    halos transition from 20\% quiescent to 80\% quiescent central
    galaxies. The circles represent results of studies based on halo
    occupation analyses. The squares are group catalogs. The pentagon
    is the prediction of UniverseMachine. The results of our fiducial
    group catalog are given with the filled red square. For comparison
    we include the results of the no-$\chi$ catalog as well.}
\end{figure}

\begin{figure}
  \epsscale{1.25}
  \hspace{-1.5cm}
  \plotone{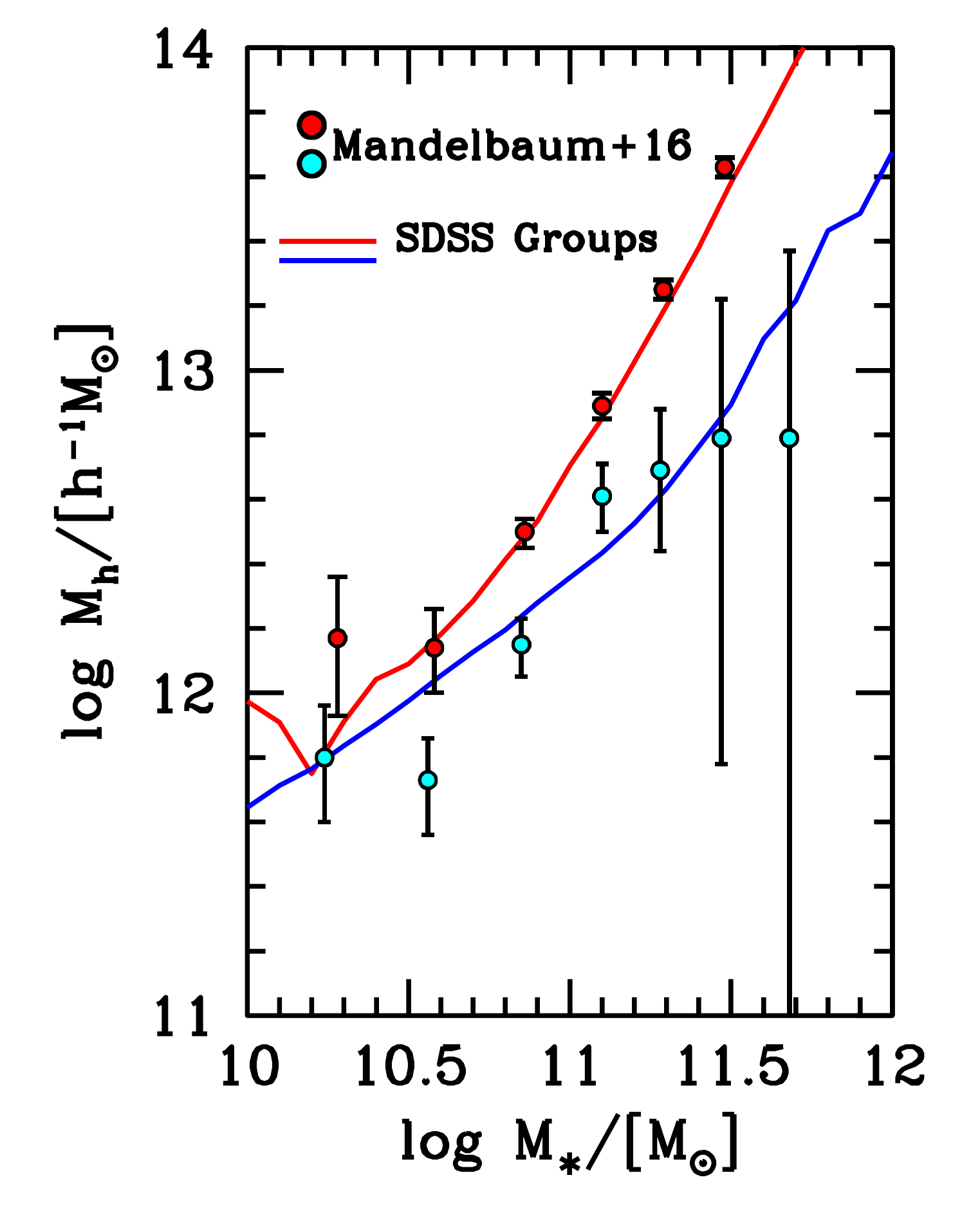}
  \vspace{-0.1cm}
  \caption{\label{f.lensing} Comparing the results of the fiducial
    group finder to the weak-lensing measurements of halo mass by
    \cite{mandelbaum_etal:16}. Red curves and symbols indicate results
    for quiescent central galaxies, while blue curves and symbols show
    results for star-forming central galaxies. }
\end{figure}

\section{Discussion}
\label{s.discussion}

The goals of this paper are two-fold: (1) How to improve the
self-calibrated approach to group finding by identifying its strengths
and deficiencies in application to real data, and (2) to better
understand how the galaxy-halo connection differs for star-forming and
quiescent samples of galaxies. We will discuss these goals in turn.

\subsection{Improving the Self-Calibration Algorithm}

The most clear failing of our current implementation of the
self-calibrated group finding algorithm is its inability to match
the clustering of faint quiescent galaxies. The primary freedom the
algorithm has to match clustering is to adjust $\bsat$ and make more
galaxies satellites of a given class. But the parameter constraints in
Figure \ref{f.params}---where $\bsatred$ is at the minimum allowed
value for all galaxies fainter than $\lgal=10^{9.5}$ $\lsolhh$
indicates that there simply aren't enough potential quiescent
satellites that exist in the actual MGS galaxy population.

The most likely reason behind this inability is that the model does
not incorporate splashback halos in the construction of the clustering
predictions. Splashback halos are host halos by our definition---they
do not exist within the virial radius of a larger halo---but have
passed through a larger halo in the past and thus are subject to
significant tidal forces (e.g., \citealt{gill_etal:05}). Treating the central galaxies
within these halos as ersatz satellite galaxies has been shown to
account for the increased quiescent fraction of central galaxies around
groups and clusters (\citealt{wetzel_etal:14}). Such galaxies increase
the overall clustering of the quiescent population because they reside
near large structures and in higher density regions.

It may be that the group finder successfully identifies these galaxies
already, but the clustering prediction is constructed by assuming that
the halo occupation functions are dependent on $\mhalo$ only. The
logical step forward is to parametrize the HOD as a function of
$\mhalo$ and $\delta_{\rm gal}$, where the second parameter is the
observed overdensity of galaxies on some scale much larger than the
typical halo radius. Theoretical predictions can be made in the same
way. Another possible second parameter is the distance to the nearest
rich group or cluster, which has been shown to have a strong impact on
many halo properties (\citealt{salcedo_etal:18}). 

The other place where the group catalog's fit of the data can be
improved is with the $\lsat(\chi)$ measurements. The results of
\cite{alpaslan_tinker:20} show no correlation between $\cgal$ and
large-scale environment for central galaxies, which is the test for
whether secondary halo properties correlate with a given galaxy
property at fixed $\mgal$. However, this test is most efficient at
detecting strong correlations---if, for example, the influence of halo
formation time on $\cgal$ was subdominant to the that of $\mhalo$ but
still weakly correlated $\cgal$, a more sensitive test would be
required.

Toward this end, including secondary halo properties in the model may
allow the group catalog to better represent the $\lsat(\chi)$
data. Additionally, this may allow the group finder to assign not just
$\mhalo$ to a group, but also constrain the formation history of the
halo that group resides in as well. To accomplish this requires new
data to constrain these extra degrees of
freedom. \cite{calderon_etal:18} have shown that marked correlation
functions can be more sensitive that other statistical methods of
detecting assembly bias is secondary galaxy properties.

In the near term, new data from the DESI-BGS will provide the
necessary statistical leverage to expand the freedom in the
self-calibration method, as well as providing enhanced constraining
power at lower galaxy luminosities where SDSS suffers from low-number
statistics. In the longer term, surveys such as WAVES
(\citealt{waves}) will provide similar space densities but to higher
redshift, allowing investigation of the time evolution of the
galaxy-halo connection in the group catalog.

\subsection{The Galaxy-Halo Connection for Star-forming and Quiescent Galaxies}

The results of the self-calibrated group finder are different from
several previous results in two main ways: the transition in halo mass
between star-forming and quiescent central galaxies, and the ratio of
the SHMRs between these two populations.

We summarize the results of Figure \ref{f.fqh}, which is quite busy,
with a summary statistic in Figure \ref{f.fqh_transition}. This
statistic is the transition width between star-forming and quiescent
halos as a function of the mass scale at which 50\% of the halos are
quiescent, which we call $M_q$.. We define the transition width as the difference in
$\log\mhalo$ between the mass scales at which $\fq$ is 0.2 and 0.8,
which we refer to as $\Delta\mhalo$. For previous studies, values of
$\log M_q$ range between 11.3 to 12.2, while the values of
$\Delta\mhalo$ range between 1.2 and 2.8\footnote{We note that
  $\Delta\mhalo$ for the CCMD results of \cite{xu_etal:18} are highly
  sensitive to the choice of 0.8 as the upper limits on $\fq$. The
  quiescent fraction rises sharply from $\fq=0.2$ to $\fq=0.7$, but
  then flattens out at higher halo masses, yielding the highest value
  of $\Delta\mhalo$. If we had chosen 0.7 as the upper limit, the
  \cite{xu_etal:18} results would have the one of the lowest values.}
The self-calibrated group finder, however, yields results in this
plane that are outside the ranges of all previous results, with the
narrowest transition region at the highest halo mass scale. We include
in this figure the group catalog results without the $\lsat(\chi)$
data for comparison. Although the results have a lower $M_q$ and
wider $\Delta\mhalo$, the changes are minimal and the results of this
catalog are still outside the range of other studies.

A primary difference between our analysis and all previous studies is
information about the halo masses for low-mass galaxies from $\lsat$
data. Although weak lensing and satellite kinematics are direct probes
of the dark matter gravitational potential around central galaxies,
the signal-to-noise of such measurements become weak at
$\mhalo\lesssim 10^{12}$ $\hmsol$. The lack of strong information on
the halos of central galaxies at these scales is evinced by the
results of Figure \ref{f.fqh_transition} and also the results Figure
\ref{f.redblue_ratio}.

Above these halo mass scales, gravitational lensing offers a
independent test of the halo masses assigned in the group
catalog. Figure \ref{f.lensing} compares the group catalog results to
the halo masses measured for SDSS central galaxies in
\cite{mandelbaum_etal:16}. To facilitate a proper comparison, the
group catalog halo masses are binned in $\mgal$, using the same MPA
stellar masses of the Mandelbaum et.~al.~analysis. The definition of
quiescent in their work is the $g-r>0.8$ color cut. This cut was shown
to be most similar to our GMM-based $\dn$ cut in Figure
\ref{f.fqcen_comp}. The bimodality in weak-lensing inferred halo
masses for the different samples is well-matched by the group catalog
results. Even though $\lsat$ is an indirect observable of $\mhalo$,
the overall amplitude of the halo mass scales of both blue and red
galaxies is reproduced in the group catalog results.

\section{Summary}

In this paper we have applied the novel self-calibrated group finding
algorithm, presented in \cite{tinker:20_nextgen}, to the full
flux-limited catalog of the SDSS Main Galaxy Sample. The group finder
categorizes each galaxy as central or satellite, and provides an
estimate of the halo mass for each group. The upgrade over previous
group-finding algorithms is that new finder has extra freedom to
assign halos to groups, and this freedom is calibrated by comparing
the predictions of the group catalog to measurements of galaxy
clustering and measurements of $\lsat$ around central galaxies.

These changes to the algorithm produce marked changes to the resulting
galaxy-halo connection inferred from group catalogs. A higher fraction
of quiescent galaxies are classified as satellites, while a lower
fraction of star-forming galaxies are satellites. At high halo masses,
star-forming central galaxies are brighter and more massive than
quiescent central galaxies in the same halos. But at low halo masses,
this ratio inverts, and quiescent central galaxies are several times
brighter and more massive than star-forming centrals.

In our catalog, the transition width between star-forming and
quiescent halos---in terms of their central galaxies---is outside the
range of previous studies, either from group catalogs or halo
occupation analyses. Our group catalog finds a narrower transition, in
terms of $\log\mhalo$, and a much higher characteristic halo mass
scale at which this transition occurs.

The application of the self-calibrated group finder to survey data has
highlighted deficiencies in the approach as well. Future
implementations will improve by incorporating possible splashback
galaxies into the inferred galaxy-halo connection, as well as
correlations between secondary halo properties and secondary galaxy
properties. New data  from the DESI-BGS will expand the
statistics that can be used in the self-calibration process, and thus allow
more freedom to properly infer the galaxy-halo connection.

The galaxy group catalog, containing central-satellite classification
and halo mass estimates for all 550,028 galaxies in the NYU-VAGC, are
available publicly\footnote{\tt https://galaxygroupfinder.net}. 

\acknowledgments

JLT gratefully acknowledges the Gauss Centre for Supercomputing
e.V. (www.gauss-centre.eu) and the Partnership for Advanced
Supercomputing in Europe (PRACE, www.prace-ri.eu) for funding the
MultiDark simulation project by providing computing time on the GCS
Supercomputer SuperMUC at Leibniz Supercomputing Centre (LRZ,
www.lrz.de). The Bolshoi simulations have been performed within the
Bolshoi project of the University of California High-Performance
AstroComputing Center (UC-HiPACC) and were run at the NASA Ames
Research Center.

\bibliography{/Users/tinker/papers/risa}
\bibliographystyle{aasjournal}


\appendix
\section{Full parameter constraints}

We use the publicly available python code {\tt emcee}
(\citealt{emcee}) to explore the posterior probability distributions
of the free parameters in the self-calibration algorithm. The
parameters of the best-fit model, as well as 68\% confidence
intervals, are listed in Table 1. In this appendix, we show corner
plots and histograms for all the parameters. The corner plots are
broke into four sets, which include the parameters that govern $\wcen$
for star-forming and quiescent galaxies, satellite threshold
parameters $\bsat$, and central weights for values of $\chi$. 

\begin{figure*}
  \epsscale{1.0}
  \hspace{-0.4cm}
  \plotone{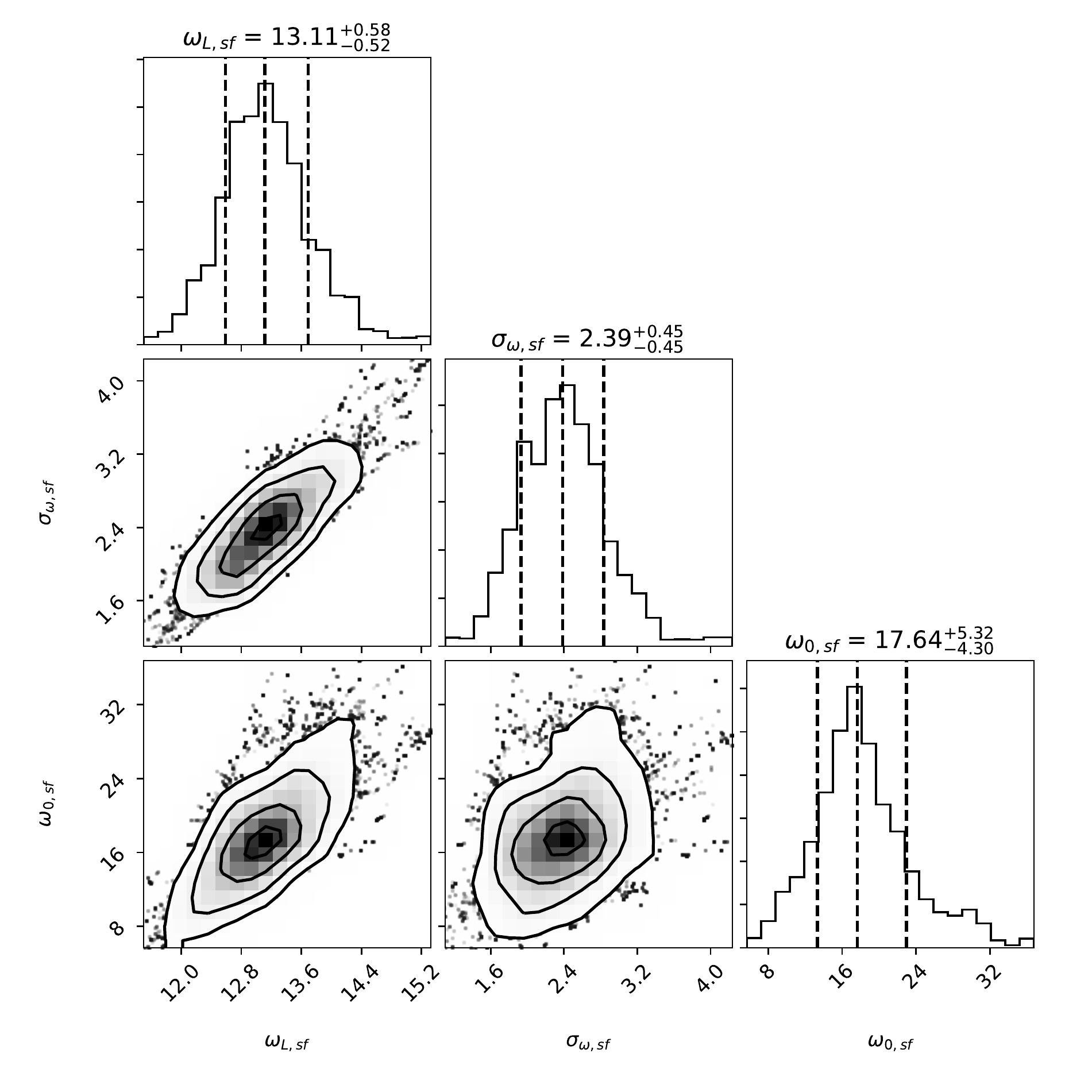}
  \vspace{-0.4cm}
  \caption{\label{f.corner1} Full parameter constraints for the
    individual parameters in the self-calibration algorithm. This
    figure shows the parameters of Eq.~(\ref{e.wcen}) for star-forming
    galaxies. Contours show 68\%, 95\%, and 99\% confidence
    regions. The histograms show the distribution of each parameter,
    with the median and 68\% confidence regions marked with vertical
    lines. }
\end{figure*}

\begin{figure*}
  \epsscale{1.0}
  \hspace{-0.4cm}
  \plotone{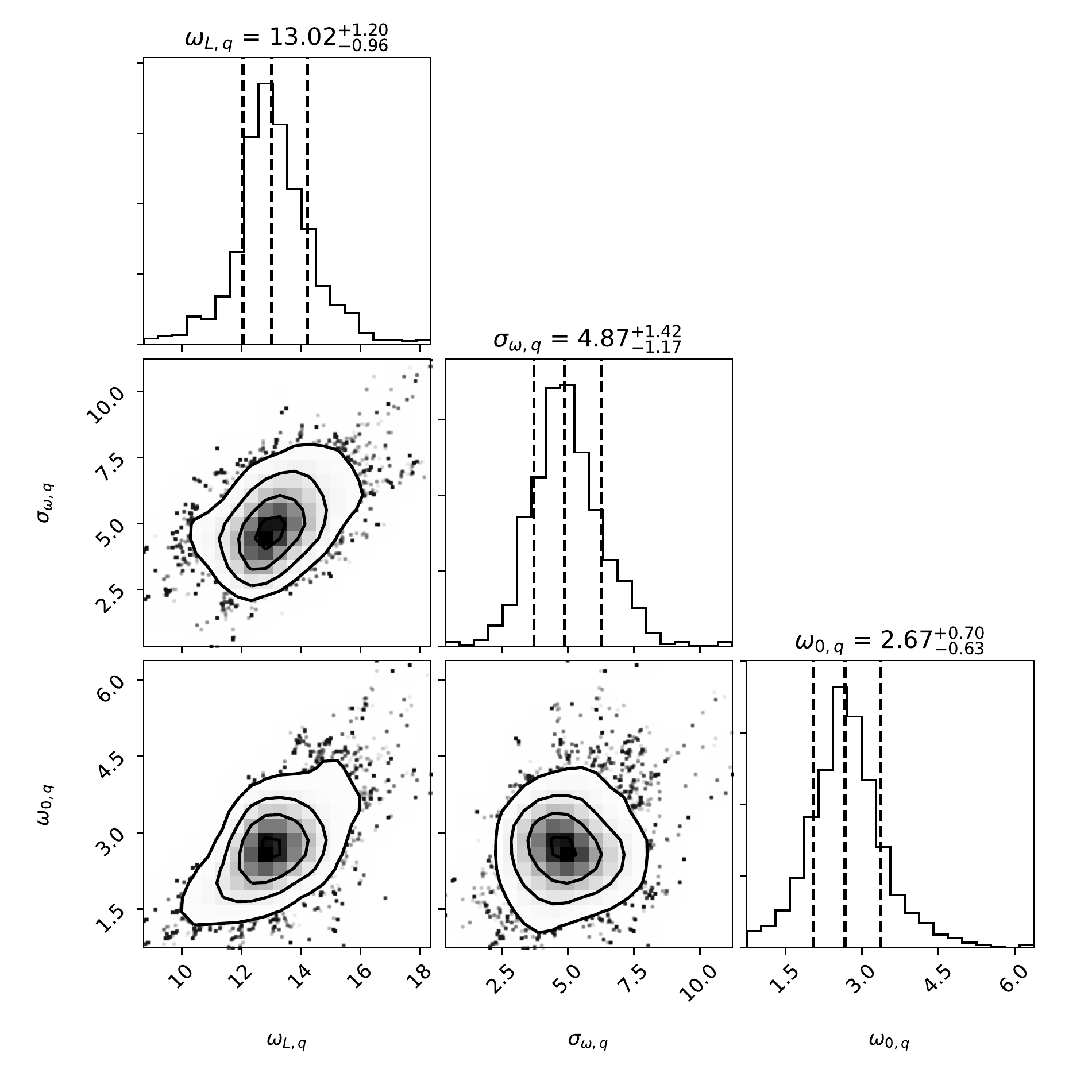}
  \vspace{-0.4cm}
  \caption{\label{f.corner2} Full parameter constraints for the
    individual parameters in the self-calibration algorithm. This
    figure shows the parameters of Eq.~(\ref{e.wcen}) for quiescent
    galaxies. Contours show 68\%, 95\%, and 99\% confidence
    regions. The histograms show the distribution of each parameter,
    with the median and 68\% confidence regions marked with vertical
    lines. }
\end{figure*}

\begin{figure*}
  \epsscale{1.0}
  \hspace{-0.4cm}
  \plotone{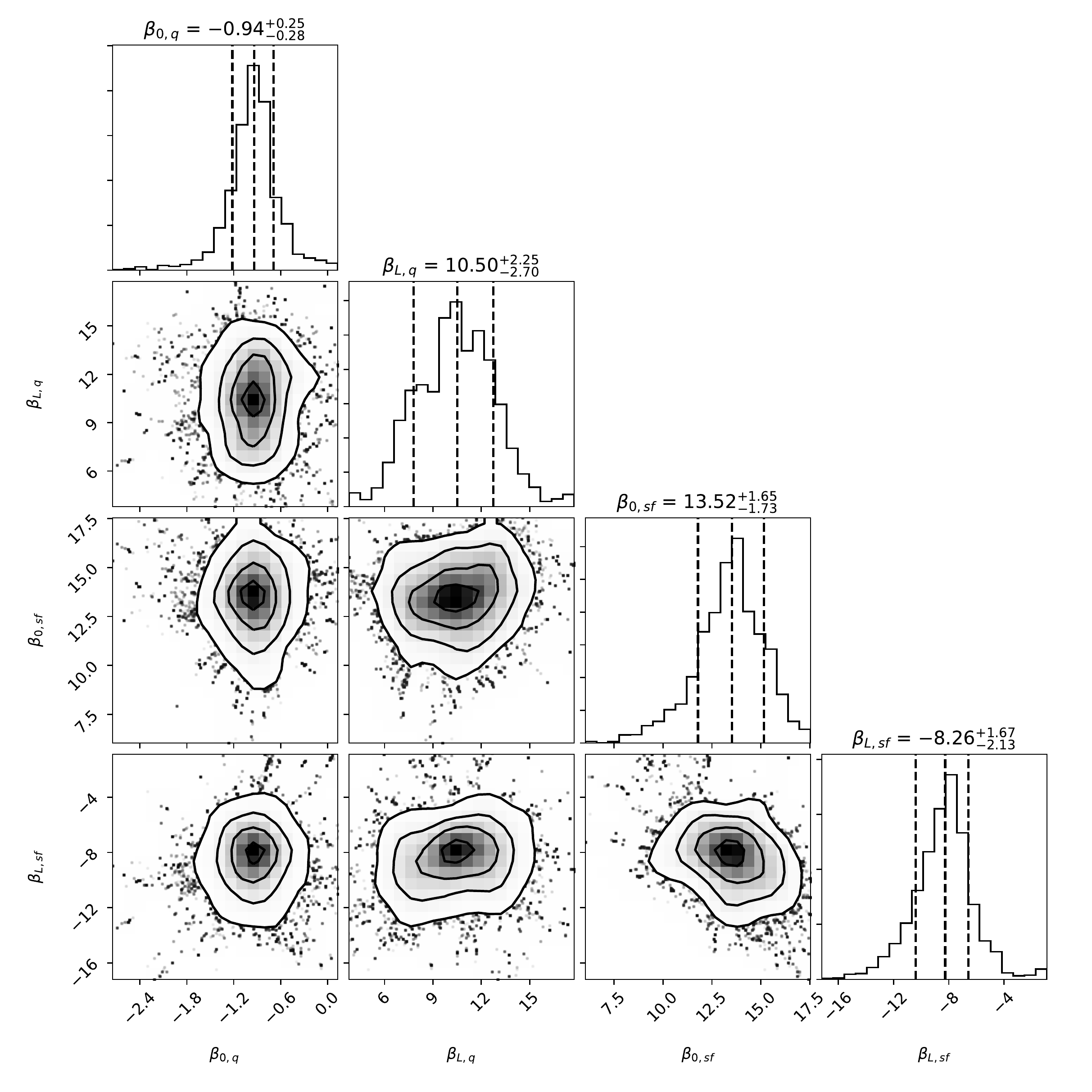}
  \vspace{-0.4cm}
  \caption{\label{f.corner3} Full parameter constraints for the
    individual parameters in the self-calibration algorithm. This
    figure shows the parameters of Eq.~(\ref{e.bsat}) for both
    quiescent and star-forming galaxies. Contours show 68\%, 95\%, and
    99\% confidence regions. The histograms show the distribution of
    each parameter, with the median and 68\% confidence regions marked
    with vertical lines. }
\end{figure*}

\begin{figure*}
  \epsscale{1.0}
  \hspace{-0.4cm}
  \plotone{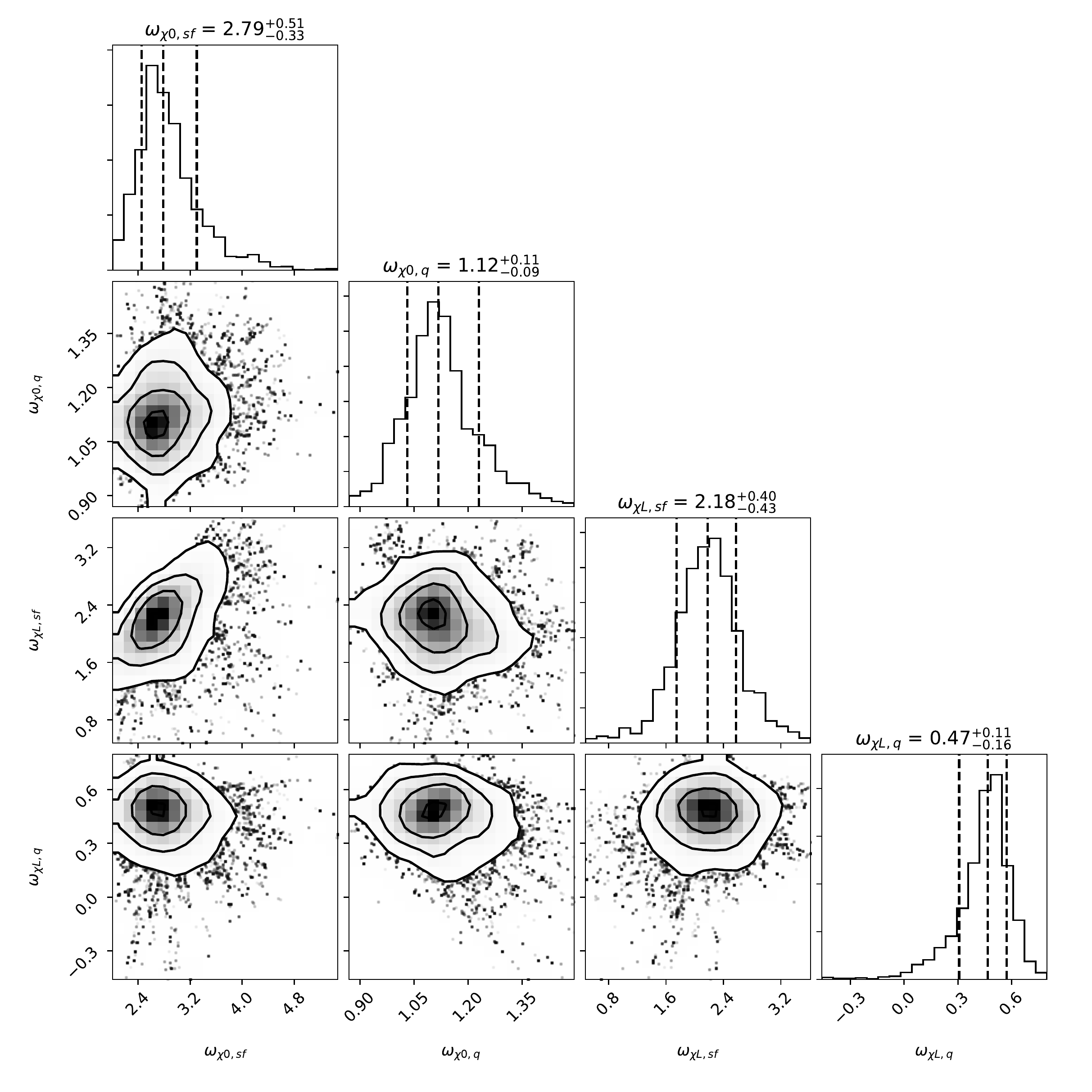}
  \vspace{-0.4cm}
  \caption{\label{f.corner4} Full parameter constraints for the
    individual parameters in the self-calibration algorithm. This
    figure shows the parameters of Eq.~(\ref{e.wchi}) for quiescent
    and star-forming galaxies. Contours show 68\%, 95\%, and 99\%
    confidence regions. The histograms show the distribution of each
    parameter, with the median and 68\% confidence regions marked with
    vertical lines. }
\end{figure*}

\end{document}